\documentclass[aps,prb,showpacs,
floatfix,
amsmath,amssymb,superscriptaddress,reprint]{revtex4-1}
\usepackage{bbold}
\usepackage{color}
\usepackage{graphicx}
\usepackage{natbib}
\usepackage{epsfig}
\usepackage{setspace}
\usepackage{amsmath}
\usepackage{amssymb}
\usepackage{verbatim}
\usepackage{tikz}
\usepackage{bibentry}

\begin{document}
\usetikzlibrary{arrows.meta,decorations.markings}
\preprint{APS/123-QED}
\title{Kerr effects in tilted multi-Weyl semimetals}
\author{Amit Gupta}
\affiliation{ Department of Physics, M. R. M. College, Lalit Narayan Mithila University, Darbhanga, Bihar 846004, India}

\begin{abstract}
We calculate the optical conductivity of type-I and type-II multi-Weyl semimetals(m-WSMs) in time-reversal(TR) broken case at finite doping. It is found that both longitudinal and Hall conductivities are renormalized by J(monopole charge) perpendicular to Weyl nodes separation and longitudal conductivity along nodes has non-trivial dependence on J. The Kerr rotations in thin films and bulk have different dependence on system parameters. The Kerr rotation in ultra-thin films is independent of monopole charge J and  depends only on Fermi velocity $v_F$ while in bulk it dependent on monopole charge J, Fermi velocity and material parameters.
\end{abstract} 

\maketitle

\section{Introduction}
The last decade has witnessed Weyl semimetals (WSMs) as a focus of research attraction due to their exotic properties. It manifest unusual Fermi arc surface states and chiral anomaly that is intimately related to topology and symmetry of WSMs \cite{armitage2018weyl,hasan2017discovery}. In WSMs, the bulk band gap closes at an even number of discrete points in the Brillouin zone(BZ). These special band closing non-generate points are referred as Weyl nodes or points \cite{wan2011topological}; Weyl nodes act as a source or sink of monopoles of Berry curvature characterized by monopole charge J. Two Weyl nodes of opposite chirality are located at opposite sign momenta when the system breaks thet time-reversal(TR symmetry; four Weyl nodes are noticed in general for systems with broken inversion(I) symmetry only \cite{armitage2018weyl,mccormick2017minimal}. Moreover, I symmetry breaking may also lead to the energy-separated Weyl points while TR symmetry breaking can result in Weyl points at the same energy \cite{dey2020dynamic,zyuzin2012weyl}. The existence of Fermi arc surface states, chiral-anomaly induced negative magnetoresistance, and the quantum anomalous Hall effect is the direct consequence of the nonzero fluxes of the Berry curvature\cite{zyuzin2012topological,son2013chiral,burkov2015chiral}. As compared to the conventional WSMs with J=1, reported in TaP, TaAs, and NbAs \cite{xu2015discovery,lv2015experimental,lv2015observation}, it has been recently shown that J can be generically greater than 1, with the crystalline environment bounding its maximum value to 3\cite{xu2011chern,yang2014classification,fang2012multi}. These are called multi-WSMs (m-WSMs). The m-WSMs can be considered as 3D analog of multilayers graphene \cite{min2008chiral}. Close to the Weyl points, m-WSMs host low-energy quasiparticles which has linear only in one direction and quadratic(cubic) in other two directions leading to usual properties which differ from linear WSMs \cite{park2017semiclassical,gorbar2017anomalous,dantas2018magnetotransport,
lepori2018axial,nag2020magneto,gupta2017floquet,gupta2019novel}. Density functional theory (DFT)calculations suggest that $HgCr_2Se_4$and $SrSi_2$can be the candidate materials for double WSM \cite{fang2012multi,xu2011chern,huang2016new} whereas $A(MoX)_3$(with A=Rb,TI; X=Te) kind of materials can accommodate triple-Weyl points\cite{liu2017predicted}.\\

Further, WSMs can be classified as type I, in this it has a conical spectrum and a point-like Fermi surface at the Weyl point which corresponds to vanishing density of states at the Weyl nodes, and type II (which occurs when  large  tilting  of  the  Weyl  cones  results  in a  Lifshitz  transition), in which an electron pocket and a hole pocket meet at a singular point of momentum space results in finite density of states, allowing for distinct topological properties \cite{armitage2018weyl,yan2017topological,xu2015structured,soluyanov2015type,
menon2018photoinduced,nag2020magneto,fei2017nontrivial,yu2016predicted,udagawa2016field,lv2017experimental}. The type-II WSM has been observed experimentally \cite{li2017evidence,kimura2019optical} while theoretical study suggests that  a  type-II  WSM  can be engineered by applying strain or chemical doping to the type-I  WSM \cite{trescher2015quantum}. The  type-II  WSM  phase is  characterized  by  a  different  class  of  Weyl  fermions manifesting the violation of Lorentz symmetry. While a lot of progress has been made experimentally and theoretically in  investigating J=1  type-I  and  type-II  WSMs,  the experimental  discovery  of  mWSMs  with $J \geq 2$  is  yet to  be  made;  however,  using  density  functional  theorycalculations   some   materials   are   conjectured   to   host Weyl  nodes  with  monopole  charges J=2,3 \cite{xu2011chern,fang2012multi,huang2016new,liu2017predicted}.\\

For thin films as well as bulk conventional WSMs, the giant Kerr rotation, polarization rotation have been demonstrated recently in the electrodynamics reponse of reflected beam, originates from the optical Hall conductivity including the quantum anomalous Hall(QAH) effect \cite{sonowal2019giant}. We extend this idea for higher monopole charge J in tilted mWSMs. We calculate analytically as well as numerically all optical conductivity tensor elements $\sigma_{ij}(\omega)$ for tilted type-I and type-II m-WSMs. 
The longitudinal conductivity along node separation has differently as a function of optical frequency than perpendicular components. This leads to different variation of Kerr rotations in ultra-thin film and bulk of the m-WSMs.
The Kerr angle has sizeable value($\approx$ 0.1 radians for $\omega< 10 ^{14}$ Hz) compared to the usually observed values in topological insulators \cite{kargarian2015theory}.\\


This paper is organized as follows: In Sec. II, We write the continuum model Hamiltonian for m-WSMs. In Sec. III, we obtain the full optical conductivity matrix for tilted type-I and type-II mWSMs, including the QAH contribution. In Sec. IV we discuss the electromagnetic response in thin films of tilted mWSM, and demonstrate the existence of giant polarization rotation is independent of monopole charge J. This is followed by a discussion of axion electrodynamics induced polarization rotation in bulk mWSMs in Sec. V and show that this rotation is dependent on monopole charge J  as well as material parameters. Finally, we conclude our main findings in Sec. VI.

\section{Model Hamiltonian}

 The non-interacting low energy effective Hamiltonian for tilted multi-Weyl semimetals is given by \cite{mukherjee2018doping,nag2020thermoelectric,nag2020magneto,menon2020anomalous},
\begin{equation}
\label{eq:ham}
\mathcal{H}_J=\xi\hbar v_tk_z\mathbb{1}_{2\times2} + \alpha_J\hbar[( \hat{k}_{-})^J \sigma_{+}+ (\hat{k}_{+})^J\sigma _{-}]+\xi\hbar v_z\hat{k}_z\sigma _z,
\end{equation}
where $\sigma_{\pm}=\frac{1}{2}(\sigma_{x}\pm i\sigma_{y})$ and $\hat{k}_{\pm}=\hat{k}_x\pm i \hat{k}_{y}$,  J represents monopole charge, $v_z$ is the effective velocity along $\hat{z}$ direction and  $\alpha_{J}$ is the material dependent parameter, e.g. $\alpha_1$ and $\alpha_2$ are the Fermi velocity and inverse of the mass respectively for the isotropic and double WSMs. \\
To avoid difficulties with anisotropic dispersions, we consider the following coordinate transformation \cite{ahn2017optical}
\begin{eqnarray}
k_x\rightarrow \rho^{1/J}\cos\phi,
k_y\rightarrow \rho^{1/J}\sin\phi,
k_z \rightarrow k_z
\end{eqnarray}
whose Jacobian is given by
\begin{equation}
\mathcal{J}=\frac{\rho^{\frac{2}{J}-1}}{J}\equiv \mathcal{J}(\rho)    
\end{equation}
In the transformed coordinates, the energy dispersion is given by $
E_{s}(\mathbf{k})=\hbar v_tk_z+s\hbar\sqrt{\alpha_n^2 (\rho_{J})^{2}+\left( k_z v_z \right)^2}$
with $\rho_J=(k_x^2+k_y^2)^{J/2}$.
 
\section{Optical Conductivity}
The interband optical conductivity in the non-interacting limit $\gamma \rightarrow 0 $ can be expressed as
\begin{eqnarray}
\sigma_{\alpha \beta}(\omega)=-\lim_{\gamma \to 0}\sum_{\bf k}\frac{i n_k}{\hbar \omega_k}\frac{M_i^{ss'}(k)M_j^{s's}(k)}{\hbar \omega+ \hbar \omega_k +i \gamma}
\end{eqnarray}

where $i,j=x,y,z, n_k^{eq}= \Theta(\mu-\hbar v_t k_z-\epsilon_J)-\Theta(\mu-\hbar v_t k_z+\epsilon_J)$ and $\hbar \omega_k=2\hbar \epsilon_j(k)$ is the energy difference between the conduction and the valence band, which is independent of the tilt velocity. The real and the imaginary parts can be separated in the limit $\gamma \rightarrow 0$, using the Sokhotski-Plemelj Formula
\begin{equation}
\lim_{\gamma \to 0}\frac{1}{x \pm i \gamma}=P\frac{1}{x} \mp i \pi \delta(x)
\end{equation}
Here P denotes the principal value.

The optical matrix element is given by
\begin{eqnarray}
M_x&=&e\alpha_J(i \sin{J \phi}-\xi \frac{v_F k_z}{\epsilon_J}\cos{J\phi}),\nonumber\\
M_y&=&e\alpha_J(-i \cos{J \phi}-\xi \frac{v_F k_z}{\epsilon_J}\sin{J\phi})\nonumber\\
M_z&=&\xi e v_F \frac{\alpha_J \rho}{\epsilon_J}
\end{eqnarray}
Here we have defined $\epsilon_J=\hbar \sqrt{\alpha_J^2 \rho^2+v_F^2 k_z^2}$. Note that $\bf{M_k}$does not depend on the tilt velocity $v_t$ at all.

\subsection{Re[$\sigma_{xx}(\omega)$]}
The real part of longitudinal optical conductivity is given by 
\begin{eqnarray}
Re[\sigma_{ii}]=-\frac{1}{(2\pi)^3}\int_{0}^{\infty} \mathcal{J} d\rho \int_{k_c}^{k_c} dk_z \int_{0}^{2\pi}\frac{n_{k}^{eq}}{\hbar \omega_k} \nonumber\\
{\mid Mi \mid }^2 \pi \delta(\omega-\omega_k)
\end{eqnarray}

The matrix element of $M_i^{ss'}(\rho, k_z) = \langle s, \rho, k_z\vert \frac{\partial\hat{H}}{\partial k_i}\vert s', \rho, k_z\rangle $ are given by
\begin{eqnarray}
M_x^{+-}(\rho, k_z)&=&J e \alpha_J \rho^{(J-1)/J}(i \sin{J \phi}-\xi \frac{v_f k_z}{\epsilon_J}\cos{J\phi}),\nonumber\\
M_y^{+-}(\rho, k_z)&=&J e \alpha_J \rho^{(J-1)/J}\alpha_J(-i \cos{J \phi}-\xi \frac{v_f k_z}{\epsilon_J}\sin{J\phi}),\nonumber\\
M_z^{+-}(\rho, k_z)&=&\xi e v_F \frac{\alpha_J \rho}{\epsilon_J}
\end{eqnarray}

Here, the optical matrix is dependent on $\phi_k$ and the angular integration leads to
\begin{eqnarray}
\int_{0}^{2\pi}{\mid Mx \mid }^2 d\phi_k=J^2\pi e^2\alpha_J^2\rho^\frac{2(J-1)}{J}\Bigl(1+\frac{v_f^2 k_z^2}{\epsilon_J^2}\Bigl)
\end{eqnarray}

\begin{figure} [t] 
\includegraphics[scale=.2]{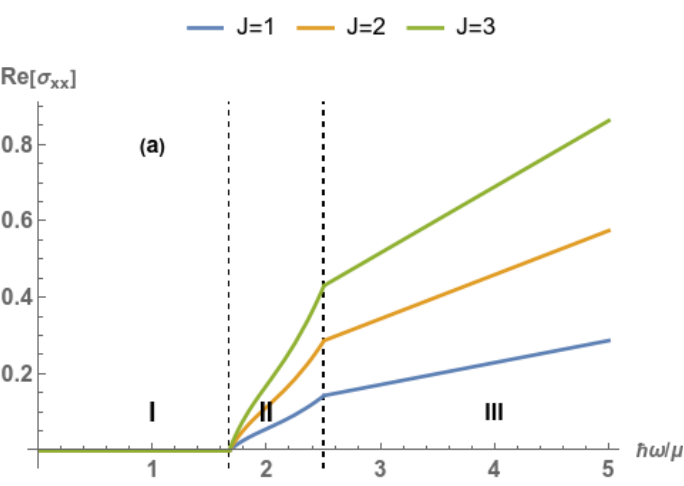} 
\includegraphics[scale=.2]{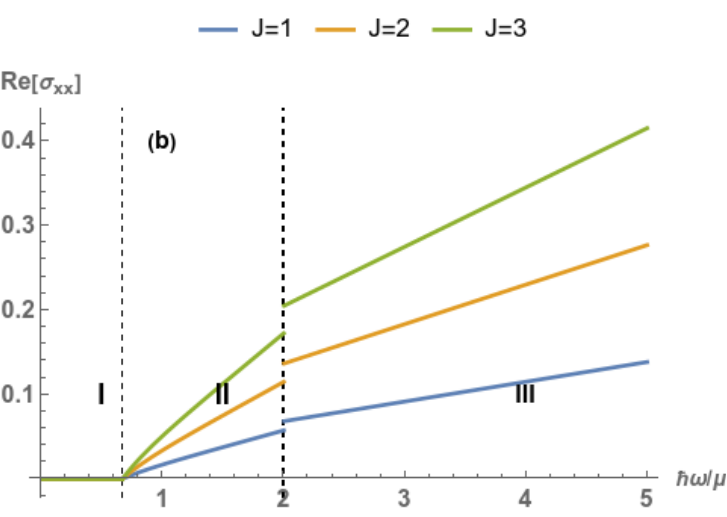}\quad
\includegraphics[scale=.2]{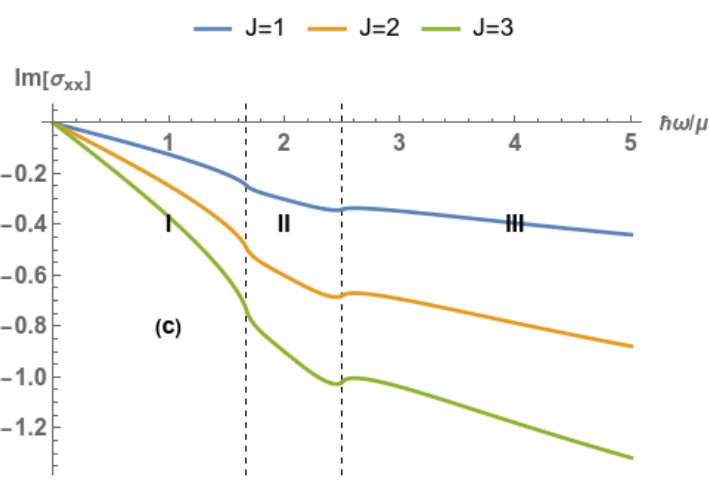}
\includegraphics[scale=.2]{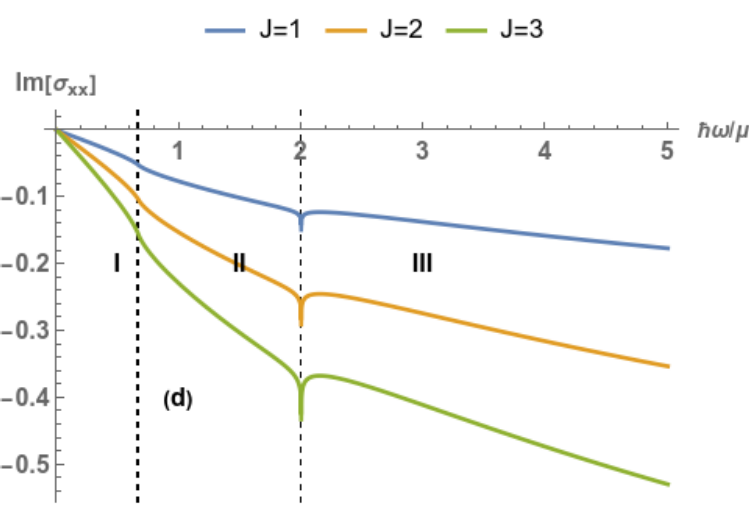}
\caption{(a),(b) show the real part of the $\sigma_{xx}$ , in a type-I ($\alpha_t = 0.2$) and a type-II ($\alpha_t  = 2.0$) mWSM, respectively.(c),(d)show the imaginary parts of the $\sigma_{xx}$ , in a type-I ($\alpha_t = 0.2$) and a type-II ($\alpha_t  = 2.0$) mWSMs, respectively.}
\label{fig_condxx}
\end{figure}
Thus we obtain
\begin{eqnarray}
Re[\sigma_{xx}(\omega)]&=&-\frac{J e^2 \alpha_J^2}{(8\pi}\int_{0}^{\infty} d\rho \rho \int_{-k_c}^{k_c} dk_z \nonumber\\
&&\frac{ \delta(\omega-\omega_k)}{\hbar\epsilon_J}\Bigl(1+\frac{v_f^2 k_z^2}{\epsilon_J^2}\Bigl)[\Theta(\mu-\hbar v_t k_z-\hbar\epsilon_J)\nonumber\\
&&-\Theta(\mu-\hbar v_t k_z+\hbar\epsilon_J)] 
\end{eqnarray}
Now, the integration for $\rho$ can easily be carried out using the root ($\rho_i$) of the argument in the Dirac delta function
\begin{equation}
\delta(f(\rho))=\sum_{x_i}\frac{\delta(\rho-\rho_i)}{\mid f'(\rho_i)\mid},\ where \ f(\rho_i)=0
\end{equation}
Substituting the roots in the delta function leads to the following one-dimensional integral over $k_z$
\begin{eqnarray}
Re[\sigma_{xx}(\omega)]&=&\frac{J e^2 }{32\pi \hbar} \int_{k_c}^{k_c} dk_z 
\Bigl[1+\Bigl(\frac{2 v_f k_z}{\omega}\Bigl)^2\Bigl]\nonumber\\
&&[\Theta(\mu-\hbar v_t k_z+\hbar \omega /2)
\nonumber\\
&&-\Theta(\mu-\hbar v_t k_z-\hbar \omega /2)] \label{cond}
\end{eqnarray}

Equation \ref{cond} can be expressed in a dimesionless form:
\begin{equation}
Re[\sigma_{xx}(\omega)]=\sigma_{\omega}^J F_{xx}, \ where \ \sigma_{\omega}^J\equiv \frac{Je^2 \omega}{6 h v_F},  
\end{equation}
and 
\begin{equation}
F_{xx}=\frac{3}{8}\int_{-1}^1 dx (1+x^2)(\Theta_{+}-\Theta_{-})\label{integral_x}   
\end{equation}

where we have defined $\Theta_{\pm}=\Theta[\frac{2 \mu}{\hbar \omega}-\alpha_t x\pm1]$, and $x=2 v_F k_z/\omega$. Evaluating the integral in Eq.(\ref{integral_x})
for the two cases of $\mid \alpha_t \mid <1$ and $\mid \alpha_t \mid >1$ leads to.

For a type-I WSM hosting a pair of oppositely tilted Weyl points with $\mid \alpha_t \mid <1$, we obtain

\begin{equation}
 Re[\sigma_{xx}(\omega)] =
    \begin{cases}
      0 & I:\omega<\omega_l,\\
      \sigma_{\omega}^J(1/2-\eta_l) & II:\omega_l<\omega<\omega_u\\
      \sigma_{\omega}^J, &  III:\omega>\omega_u
    \end{cases}  
\end{equation}
 where we have defined conductivity via optical frequency:$\sigma_{\omega}=e^2\omega/(6 h v_f)$. The photon energy bounds are $\hbar \omega
 _l=2\mu/(1+\mid \alpha_t \mid)$,$\hbar \omega_u =2\mu/(1-\mid \alpha_t \mid)$ and
 \begin{equation}
\eta_l=\frac{3}{8\mid \alpha_t \mid}\Bigl(\frac{2\mu}{\hbar\omega}-1\Bigl)\Bigl[1+\frac{1}{3\alpha_t^2}(\frac{2\mu}{\hbar\omega}-1)^2\Bigl]     
 \end{equation}

In region I, for $\omega<\omega_l$, the optical transitions is absent due to Pauli blocking. In region II, $\omega \in (\omega_l,\omega_u) $, optical transitions happen resulting in finite $Re[\sigma_{xx}(\omega)]$. In the $\alpha_t \rightarrow 0$ limit we have $\omega_l \rightarrow \omega_u= 2\mu/\hbar $ and we have $Re[\sigma_{xx}(\omega)]=\sigma_{\omega}$ for $\hbar \omega \geq 2\mu$. In the other limit $\alpha_t \rightarrow 1$, we have $\omega_u\rightarrow\infty$ with region II extending to higher energies.
\\
For a type-II WSM with $\mid \alpha_t \mid >1$, we obtain
\begin{equation}
 Re[\sigma_{xx}(\omega)] =
    \begin{cases}
      0 & I:\omega<\omega_l,\\
      \sigma_{\omega}^J(1/2-\eta_l) & II:\omega_l<\omega<\omega_u\\
      \sigma_{\omega}^J\eta_2, &  III:\omega>\omega_u
    \end{cases}  
\end{equation}
Here we have defined $\hbar \omega_u =2\mu/(\mid \alpha_t \mid-1)$ and
\begin{equation}
\eta_2=\frac{3}{4 \mid \alpha_t\mid}\Bigl[1+\frac{1}{3\alpha_t^2}+\Bigl(\frac{2\mu}{\alpha_t \hbar\omega}\Bigl)^2\Bigl]    
\end{equation}

The real and imaginary parts of conductivity are related by Kramers-kronig relations
\begin{eqnarray}
Im[\sigma_{ij}(\omega)]=-\frac{2\omega}{\pi}\mathcal{P}\int_0^{\Lambda}d\omega'\frac{Re[\sigma_{ij}(\omega')}{\omega'^2-\omega^2}\nonumber\\
Re[\sigma_{ij}(\omega)]=\frac{2}{\pi}\mathcal{P}\int_0^{\Lambda}d\omega'\frac{\omega' Im[\sigma_{ij}(\omega')}{\omega'^2-\omega^2}
\label{KK_relations}
\end{eqnarray}

where $\mathcal{P}$ refers to the principle value of the integral and $\Lambda(>>\omega)$ is the ultravoilet cutoff. \\
The imaginary part of the longitudinal optical conductivity $\sigma_{xx}(\omega)$ can be calculated analytically \cite{sonowal2019giant} using above Eq.(\ref{KK_relations}) and plotted in Fig.(\ref{fig_condxx}).

\begin{figure} [t] 
\includegraphics[scale=.2]{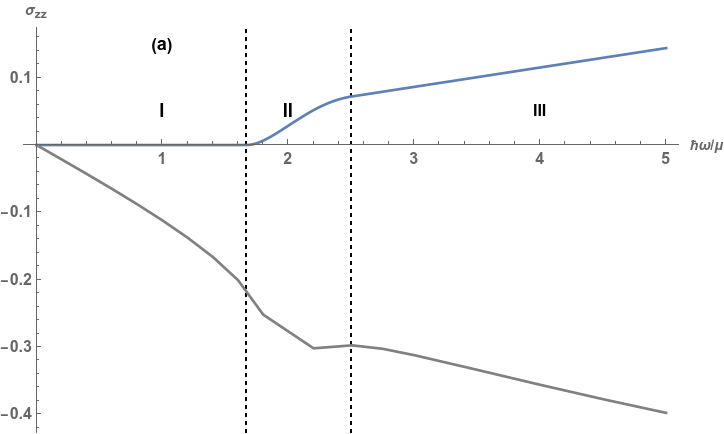} 
\includegraphics[scale=.2]{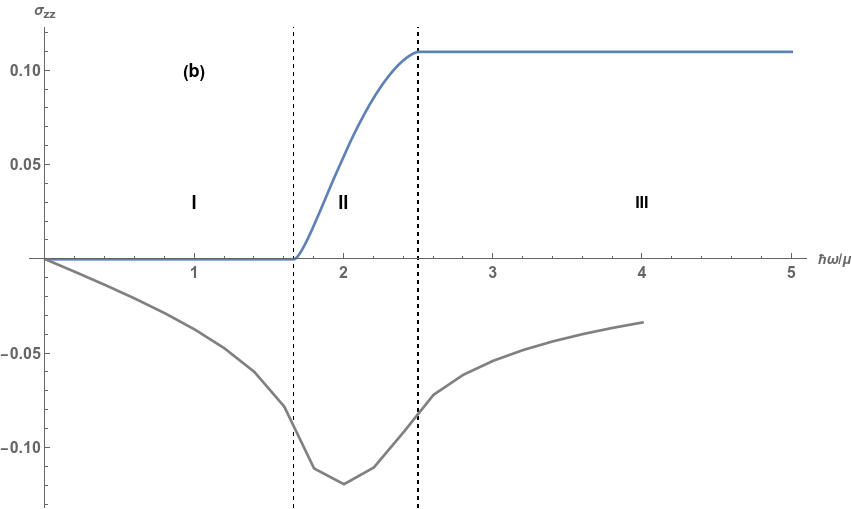}\quad 
\includegraphics[scale=.25]{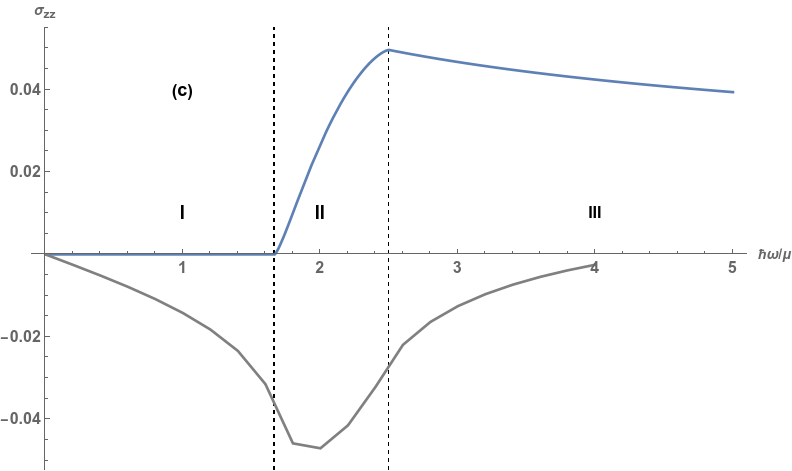}
\caption{(a),(b),(c) show the real(blue) and the imaginary(grey) parts of the $\sigma_{zz}$ , in a type-I ($\alpha_t = 0.2$) mWSM respectively. Here we have chosen $\mu=.125$ eV, $v_F=10^6 m/s$, $Q=3.25\times 10^8 m^{-1}$, $\alpha_2=.009  m^2/s$, $\alpha_3=4.5 \times 10^{-11} m^3/s$} 
\label{fig_condzz}
\end{figure}

\subsection{Re[$\sigma_{zz}(\omega)$]}

Following the same procedure a s above, the angular integration leads to

\begin{eqnarray}
\int_{0}^{2\pi}{\mid Mz \mid }^2 d\phi_k=2\pi e^2 v_f^2 \Bigl(\frac{\alpha_J^2 \rho^2}{\alpha_J^2 \rho^2+v_F^2 k_z^2}\Bigl)
\end{eqnarray}
Thus the longitudinal optical conductivity in this case is given by
\begin{eqnarray}
Re[\sigma_{zz}(\omega)]&=&-\frac{e^2 v_f^2\alpha_J^2 }{4\pi J}\int_{0}^{\infty} d\rho  \int_{-k_c}^{k_c} dk_z \frac{\rho^{\frac{2}{J}+1}}{\alpha_J^2 \rho^2+v_F^2 k_z^2}\nonumber\\
&&\frac{ \delta(\omega-\omega_k)}{\hbar\epsilon_J}[\Theta(\mu-\hbar v_t k_z-\hbar\epsilon_J)\nonumber\\
&&-\Theta(\mu-\hbar v_t k_z+\hbar\epsilon_J)] 
\end{eqnarray}

Performing the $\rho$ integral using the roots of the $\delta$ function, we have
\begin{eqnarray}
Re[\sigma_{zz}(\omega)]&=&\frac{e^2 v_F^2}{16\pi \hbar J \alpha_J^{2/J}} \int_{-\omega/2v_F}^{\omega/2v_F} dk_z 
\Bigl[1-\Bigl(\frac{2 v_F k_z}{\omega}\Bigl)^2\Bigl]\nonumber\\
&&[\Theta(\mu-\hbar v_t k_z+\hbar \omega /2)
\nonumber\\
&&-\Theta(\mu-\hbar v_t k_z-\hbar \omega /2)] \label{cond_z}
\end{eqnarray}

Eq.(\ref{cond_z}) can be expressed in a dimesionless form:
\begin{equation}
Re[\sigma_{zz}(\omega)]=\sigma_{\omega,z}^J F_{zz}, \ where \ \sigma_{\omega,z}^J\equiv \frac{e^2 (\frac{\omega}{2})^{\frac{2}{J}-1}v_f}{3 h J \alpha_J^{2/J}},  
\end{equation}
and 
\begin{equation}
F_{zz}=\frac{3}{8}\int_{-1}^1 dx (1-x^2)^{1/J}(\Theta_{+}-\Theta_{-})\label{integral_x}   
\end{equation}
where $\Theta_{\pm}\equiv\Theta(\mu-\alpha_t x \pm\hbar\omega/2)$. The imaginary part has analytical only in case of J=1\cite{sonowal2019giant}. For higher J, the imaginary parts are computed numerically. The real and imaginary parts are plotted in Fig.(\ref{fig_condzz}).

\begin{figure} [t] 
\includegraphics[scale=.18]{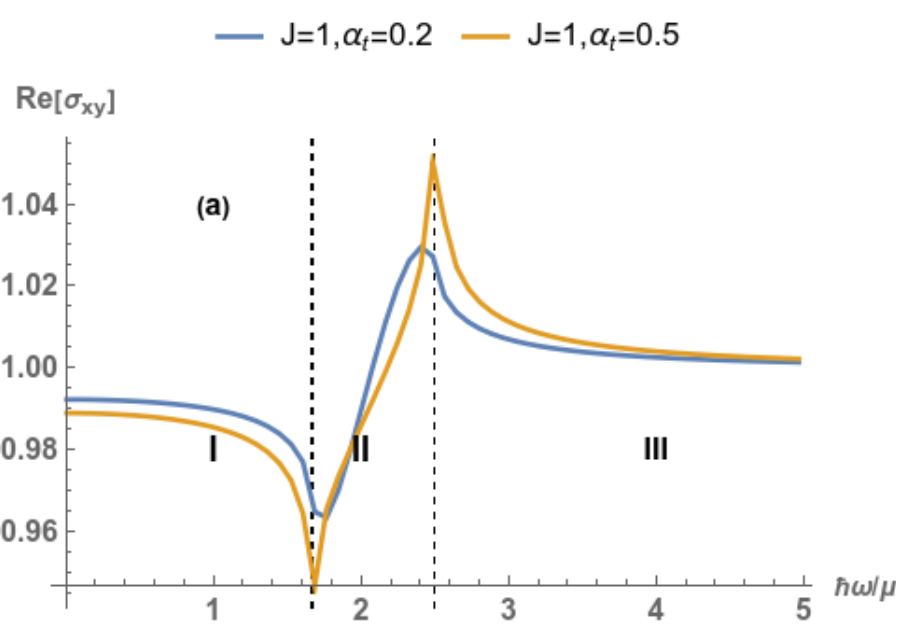} 
\includegraphics[scale=.18]{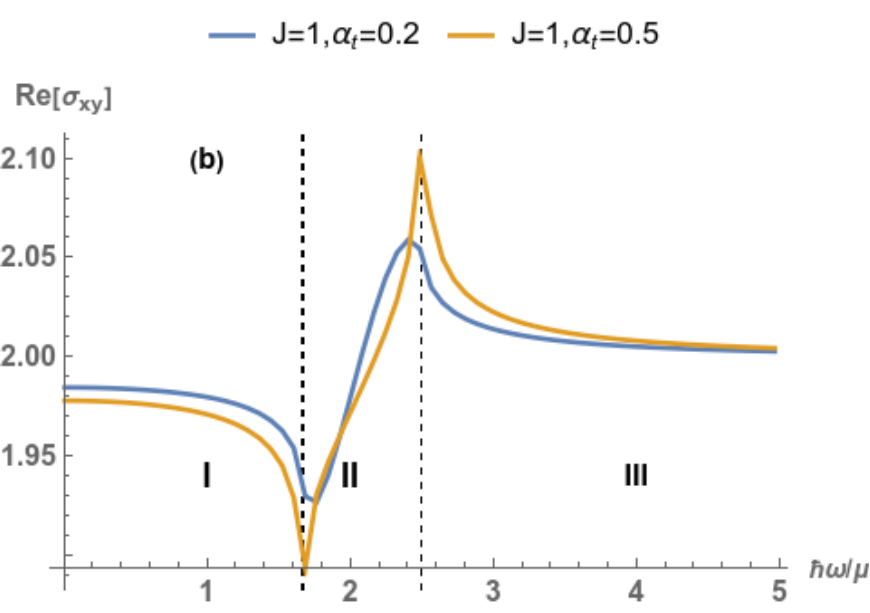}\quad 
\includegraphics[scale=.25]{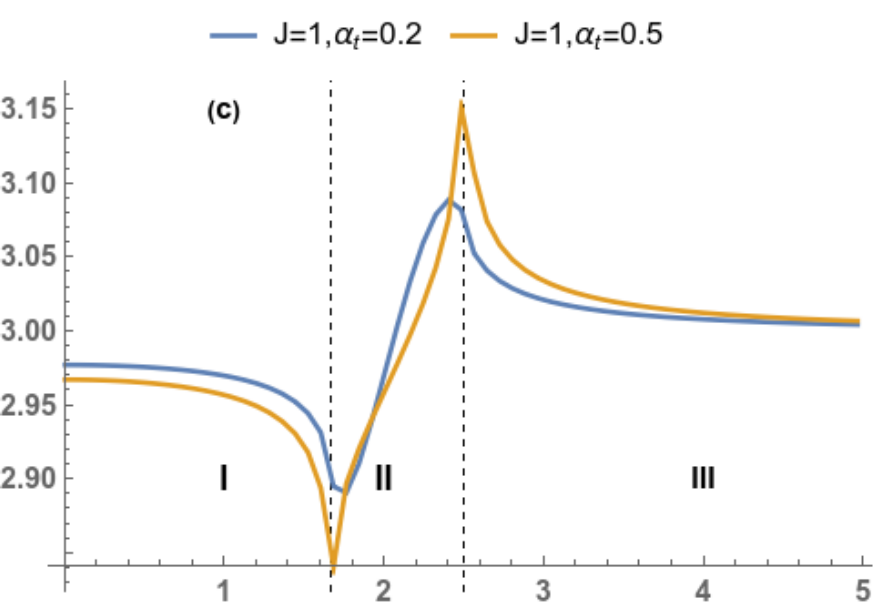}
\caption{(a),(b) and (c) show real part of  $\sigma_{xy}$ , in a type-I ($\alpha_t = 0.2,0.5 $)  mWSM, respectively.} 
\label{fig_condxy_im}
\end{figure}

\subsection{Im[$\sigma_{xy}(\omega)$]}
The transverse optical Hall conductivity is given by
\begin{eqnarray}
\sigma_{xy}(\omega)]=\frac{1}{(2\pi)^3}\int_{0}^{2\pi}d\phi_k\int_{0}^{\infty} d\rho  \int_{k_c}^{k_c} dk_z \sigma_k^{ij}(\omega)
\end{eqnarray}
where the conductivity kernel is given by
\begin{equation}
\sigma_k^{ij}=\frac{-n_k^{eq}}{\hbar \omega_k}\Bigl(\frac{M_j M_i^*}{\omega + \omega_k +i \gamma}+\frac{M_j^* M_i}{\omega - \omega_k +i \gamma}\Bigl)
\end{equation}

\begin{figure} [t] 
\includegraphics[scale=.21]{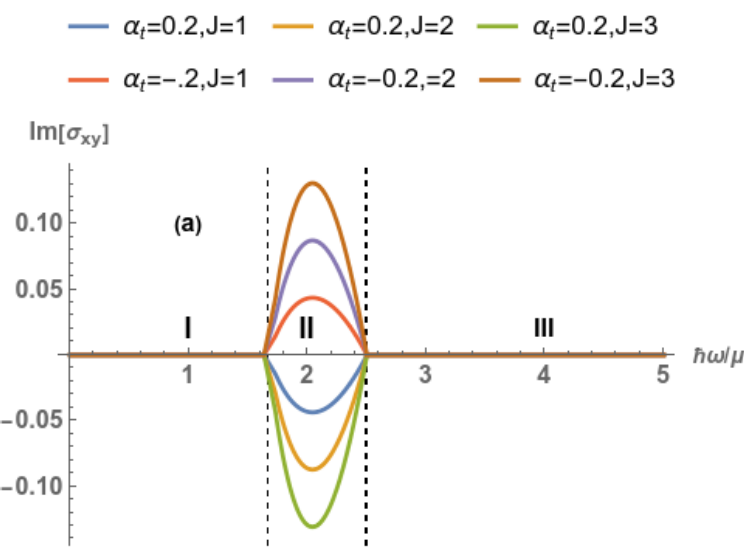} 
\includegraphics[scale=.17]{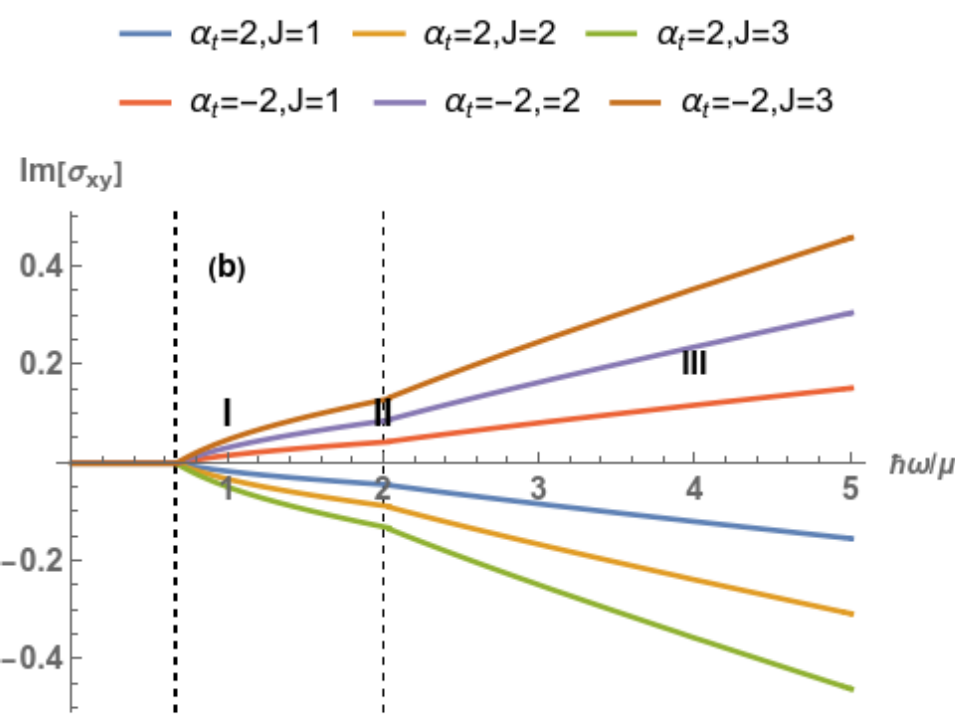}\quad
\includegraphics[scale=.25]{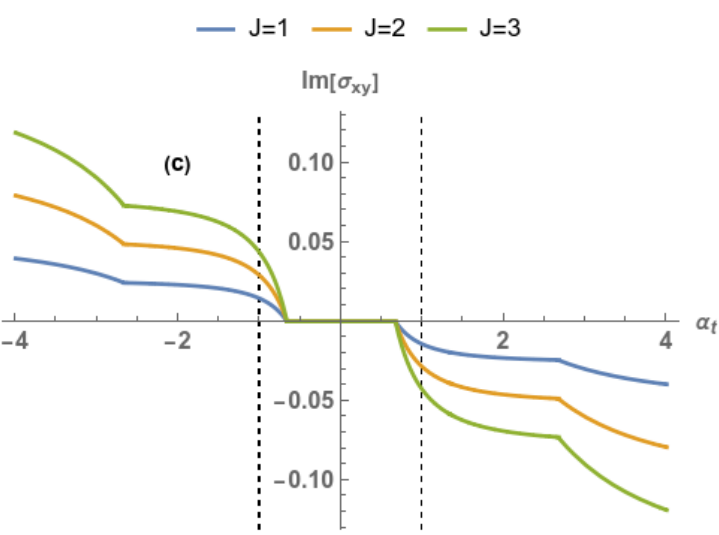}
\caption{(a),(b) show the imaginary part of $\sigma_{xy}$ , in a type-I ($\alpha_t = -0.2,0.2$) and a type-II ($\alpha_t  = -2.0,2.0$) mWSM, respectively as a function of optical frequency $\omega$.(c)show the imaginary part of  $\sigma_{xy}$ as a function of tilt parameter $\alpha_t$. } 
\label{fig_condxy_re}
\end{figure}
\begin{eqnarray}
M_y M_x^*&=&-1+\Bigl(\frac{v_f k_z}{\epsilon_j}\Bigl)^2\sin{J\phi_k}\cos{\phi_k}+i \frac{v_f k_z}{\epsilon_J}\nonumber\\
&&=M_y^*M_x
\end{eqnarray}

Performing the angular integration over $\phi_k$
\begin{equation}
\int_{0}^{2\pi}\Bigl(M_y M_x^*+M_y^*M_x\Bigl)=i\xi2\pi (\alpha_J J)^2\rho^{2(J-1)/J}\frac{v_f k_z^{\xi}}{\epsilon_J}    
\end{equation}

The imaginary part of $\sigma_{xy}(\omega)$ can be expressed as
\begin{eqnarray}
Im[\sigma_{xy}(\omega)]&=&\frac{Je^2\alpha_J^2}{4\pi }\sum_{\xi=\pm 1}\xi\int_{0}^{\infty}\rho d\rho\int_{k_c}^{k_c}dk_z^{\xi} \frac{v_F k_z^{\xi}}{\epsilon_J}\nonumber\\
&&\frac{n_k^{eq}}{\hbar\omega_k}\delta(\omega-\omega_k)   
\end{eqnarray}

Performing the $\rho$ integral and the $\delta$ function lead to the following one-dimensional integral
\begin{eqnarray}
Im[\sigma_{xy}(\omega)]&=&\frac{J e^2v_F}{8\pi \hbar \omega}\sum_{\xi=\pm 1}\xi\int_{-\omega/2v_F}^{\omega/ 2v_F }k_z^{\xi}dk_z^{\xi}\nonumber\\
&&\Bigl[\Theta\Bigl(1-\frac{\hbar \omega}{2 \mu}-\frac{\hbar \xi v_t k_z^{\xi}}{\mu}\Bigl)\nonumber\\
&&-\Theta\Bigl(1+\frac{\hbar \omega}{2 \mu}-\frac{\hbar \xi v_t k_z^{\xi}}{\mu}\Bigl)\Bigl]
\label{hall_cond}
\end{eqnarray}

\begin{figure} [t] 
\includegraphics[scale=.2]{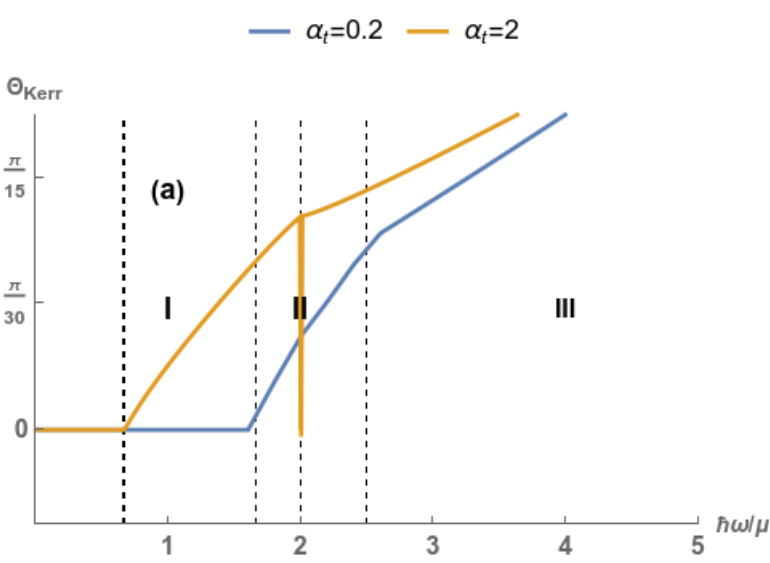} 
\includegraphics[scale=.16]{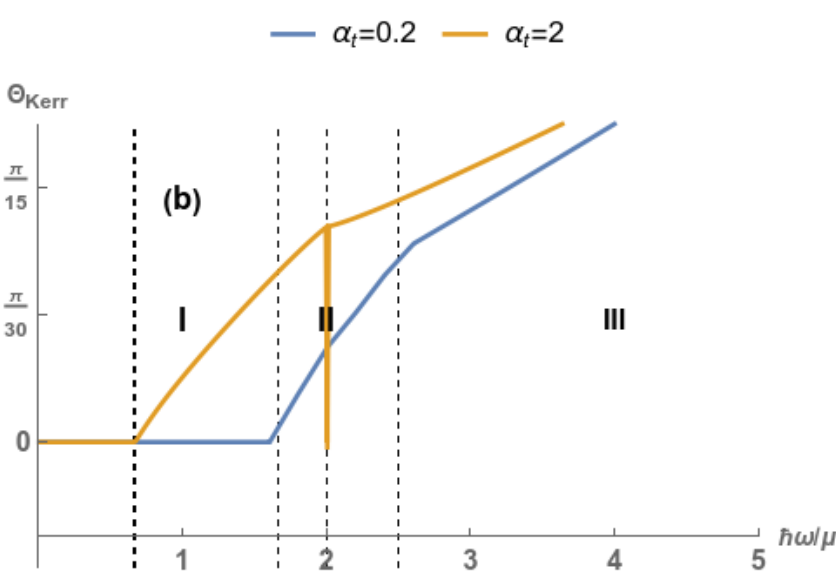}\quad 
\includegraphics[scale=.25]{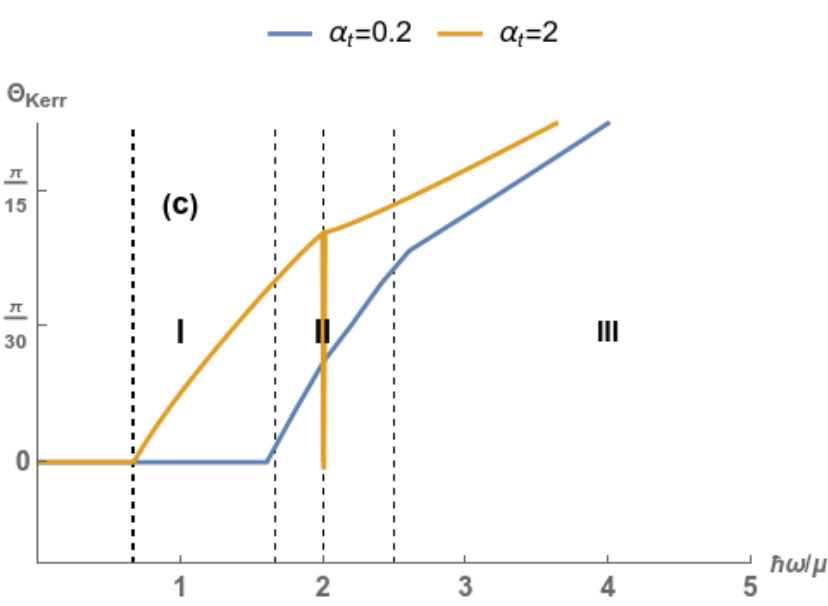}
\caption{(a),(b) and (c) show the Kerr angle as a function of optical frequency for type-1($\alpha_t = 0.2$) and a type-II ($\alpha_t  = 2.0$)for single, double and triple WSM, respectively.} 
\label{fig_kerr_angle}
\end{figure} 

\begin{figure}[t]
\includegraphics[scale=.16]{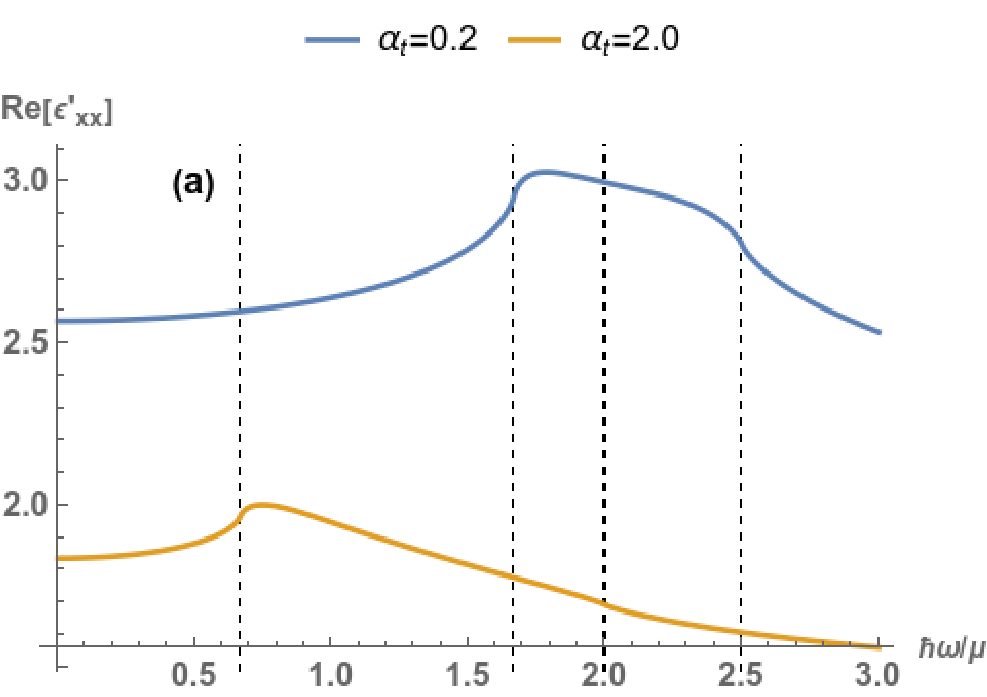} 
\includegraphics[scale=.16]{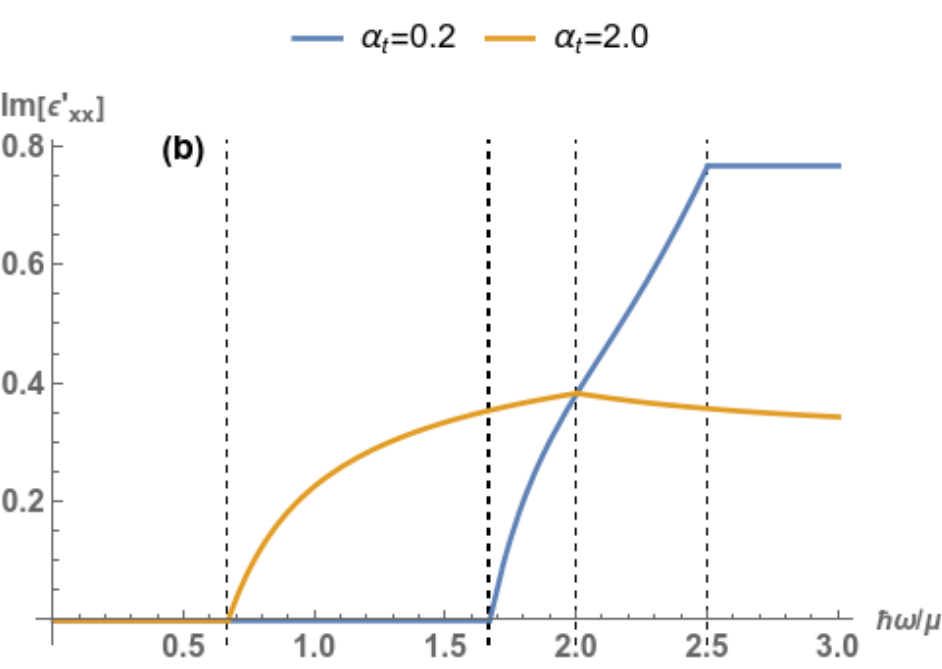} 
\includegraphics[scale=.17]{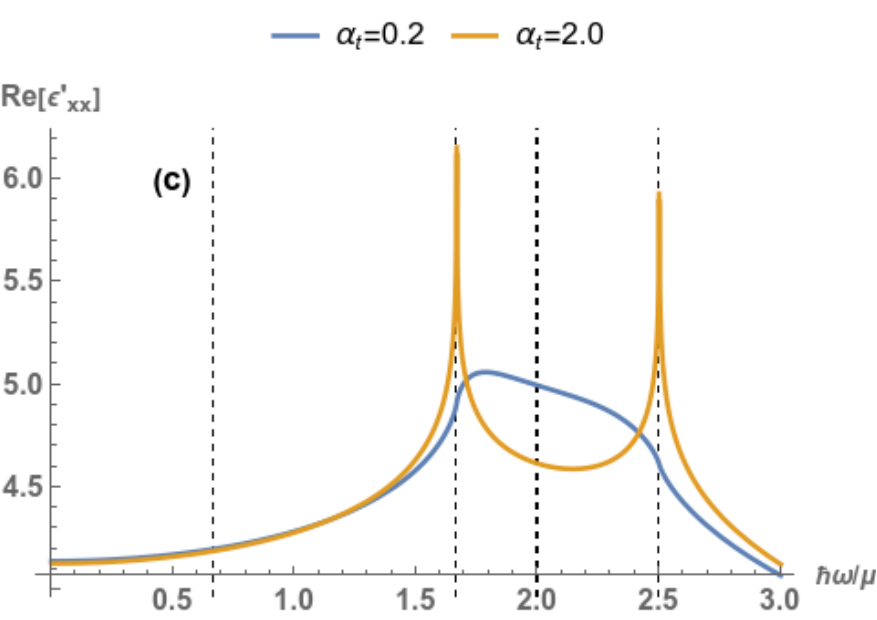}\quad
\includegraphics[scale=.17]{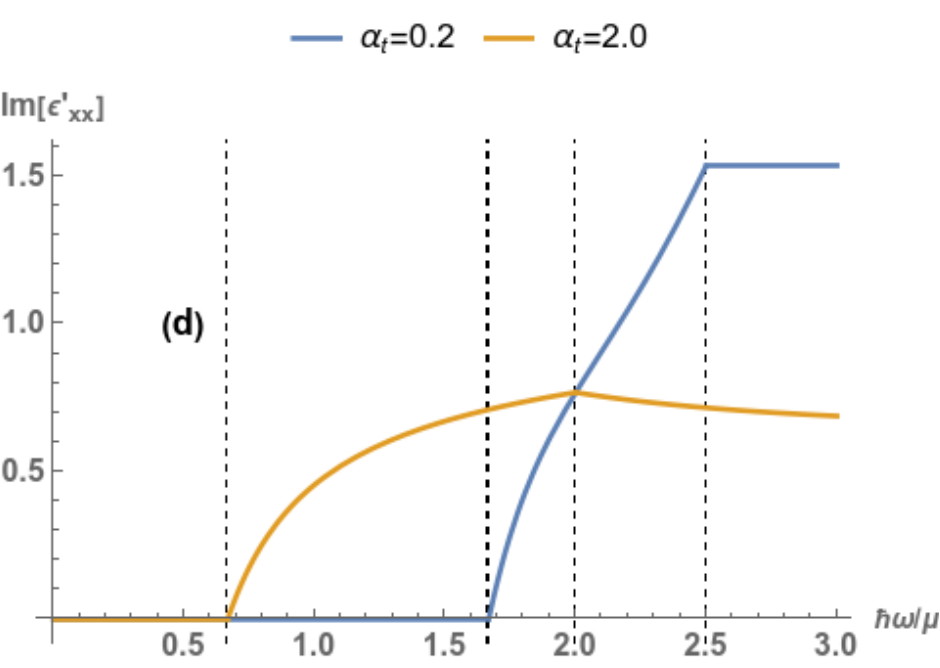}
\includegraphics[scale=.17]{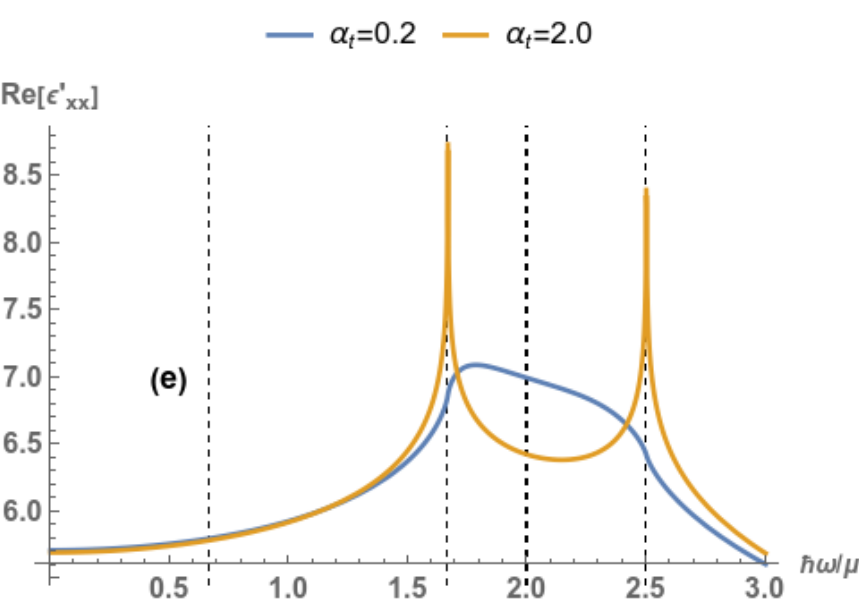}
\includegraphics[scale=.17]{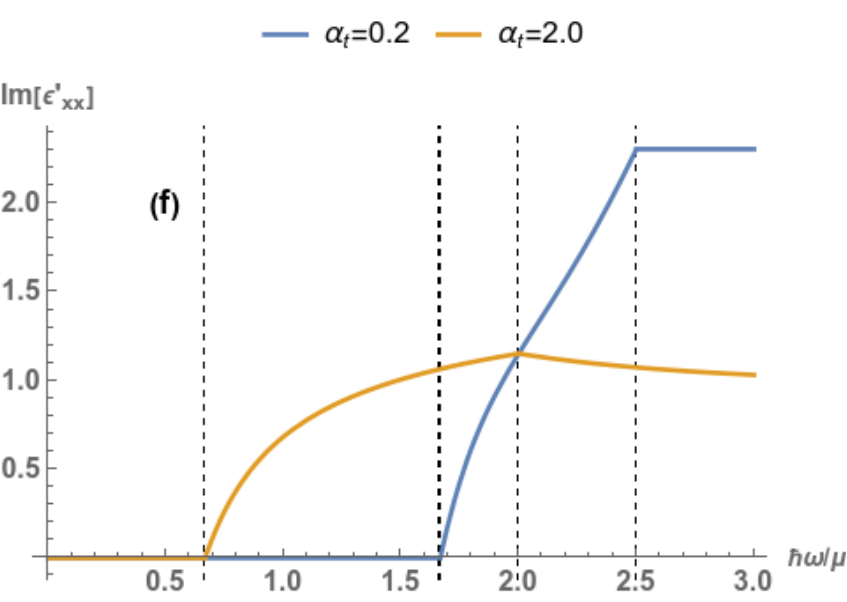}
\caption{(a),(b) show the real and imaginary parts of effective dielectric term $\varepsilon_{xx}$ for single WSM at  ($\alpha_t = 0.2,2.0 $). (c),(d) show the real and imaginary parts of effective dielectric term $\varepsilon_{xx}$ for double WSM at  ($\alpha_t = 0.2,2.0 $).(e),(f) show the real and imaginary parts of effective dielectric term $\varepsilon_{xx}$ for triple WSM at  ($\alpha_t = 0.2,2.0 $).} 
\label{fig_dielectric_xx}
\end{figure}

Eq.(\ref{hall_cond}) can be expressed as 
\begin{equation}
Im[\sigma_{xy}(\omega)]=J\frac{3\sigma_{\omega}^J}{2 }\frac{\mu^2}{\hbar^2\omega^2} F_{xy}  
\end{equation}

where we have defined 
\begin{equation}
F_{xy} =\sum_{\xi=\pm 1}\xi\int_{-\hbar\omega/2\mu}^{\hbar\omega/2\mu} x dx 
\sum_{p=\pm1}\Theta\Bigl(1-\xi \alpha_t x-p\frac{\hbar \omega}{2\mu}\Bigl)
\end{equation}

The real part of the Hall conductivity can be computed analytically using Eq.(\ref{KK_relations})\cite{sonowal2019giant}. The real part has also contribution due to Weyl nodes separation $\frac{J\sigma_Q}{\mid \alpha_t\mid}$ \cite{nag2020thermoelectric}. The real and imaginary parts of Hall conductivity are plotted in Figs.(\ref{fig_condxy_im}) and(\ref{fig_condxy_re}).

\section{Kerr rotation in thin films of m-WSMS}
In the ultra-thin limit, the thickness of the film(d) $\ll$  the wavelenth($\lambda$) of light $\lambda$, even though d$\gg$a, the lattice constant. Under these conditions the m-WSM film can be considered as a 2D surface. Thus the polarization angle rotation can be obtained simply by matching the electromagnetic boundary conditions (at z=0) on either side of the thin film.\\

The polarization angle $\Phi_{kerr}$ and the azimuth of the major axes of the polarization ellipse $\Psi_{Kerr}$. For the case of a reflected beam, the corresponding dimensionless complex quantities are defined as

\begin{figure} [t] 
\includegraphics[scale=.2]{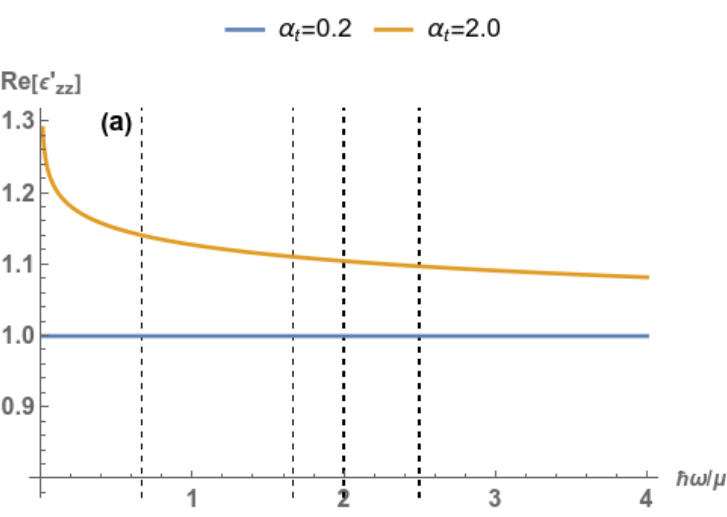} 
\includegraphics[scale=.2]{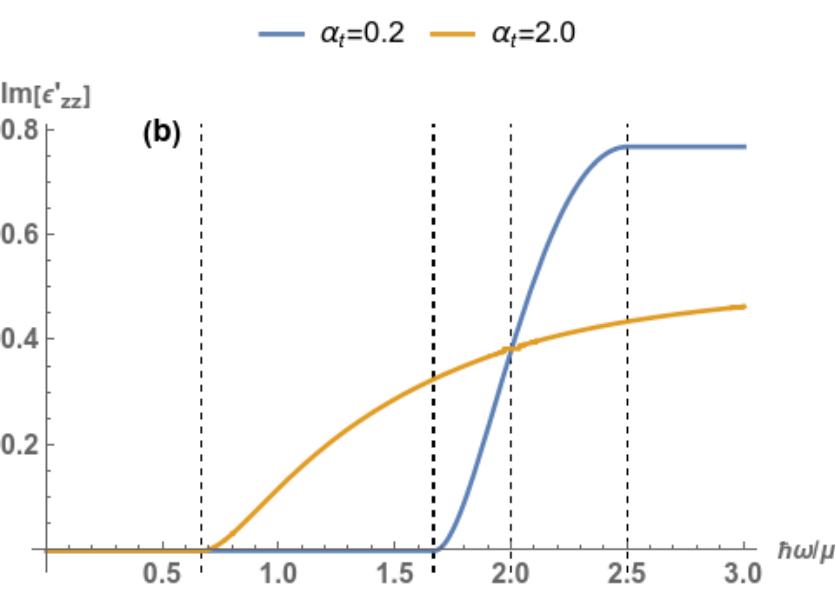} 
\includegraphics[scale=.17]{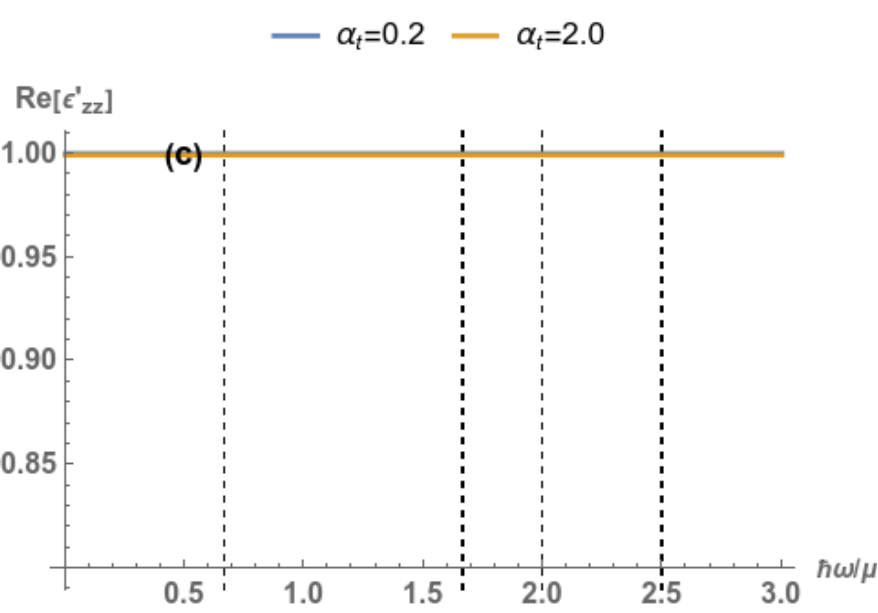}\quad
\includegraphics[scale=.17]{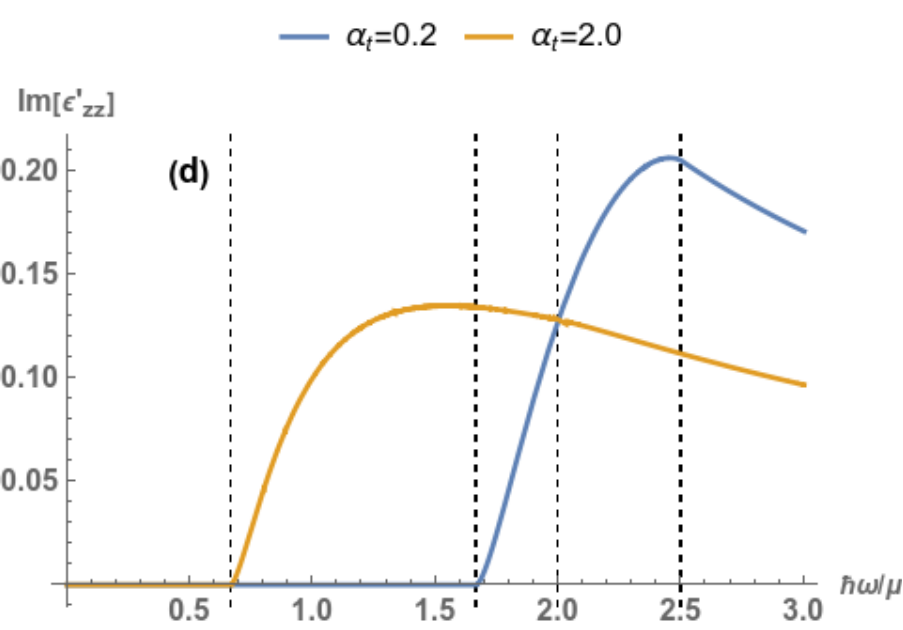}
\includegraphics[scale=.17]{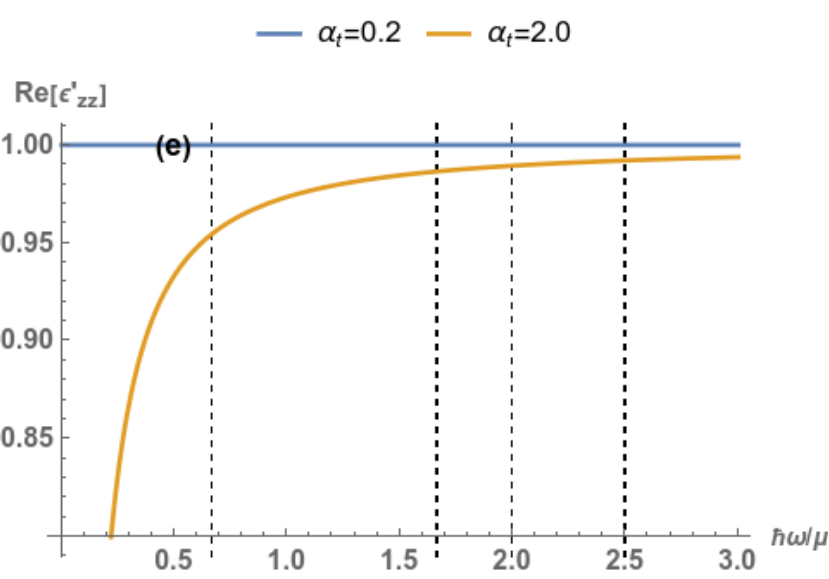}
\includegraphics[scale=.17]{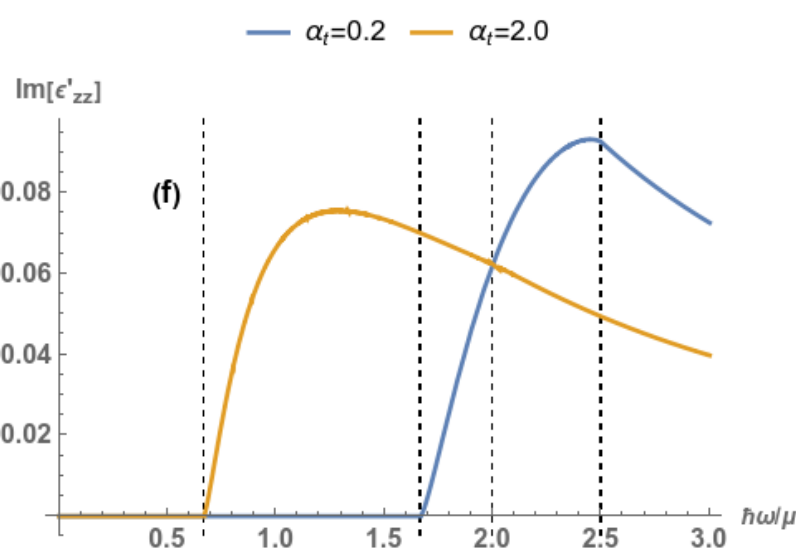}
\caption{(a),(b) show the real and imaginary parts of effective dielectric term $\varepsilon_{zz}$ for single WSM at  ($\alpha_t = 0.2,2.0 $). (c),(d) show the real and imaginary parts of effective dielectric term $\varepsilon_{zz}$ for double WSM at  ($\alpha_t = 0.2,2.0 $).(e),(f) show the real and imaginary parts of effective dielectric term $\varepsilon_{zz}$ for triple WSM at  ($\alpha_t = 0.2,2.0 $).}
\label{fig_dielectric_zz}
\end{figure}

\begin{equation}
\chi_{Kerr}^s=\frac{r_{ps}}{r_{ss}} \ and \ \chi=-\frac{r_{sp}}{r_{pp}}
\label{Kerr_angle}
\end{equation}
where $r_{ij}$ = $(s,p)$ are reflection coefficients. In TR broken m- WSM, the surface conductivity has finite off-diagonal terms for the surface without Fermi arc states(surfaces $\perp \textbf{Q}$) while it is diagonal for the surface hosting Fermi arcs states(surfaces $\parallel \textbf{Q}$). We consider these two cases in the following subsections.\\

\subsection{Incidence on surface without Fermi arc states($\perp Q$)}
Consider a linearly polarized incident beam propagates(say in direction $\hat{\textbf{z}})$ parallel to Q and the electric polarization is in the x-y plane. Let us assume that the interface of air and m-WSM thin film is located at the $z=0$ plane. The wave vectors for the incident, reflected and refracted beam are
\begin{figure} [t] 
\includegraphics[scale=.2]{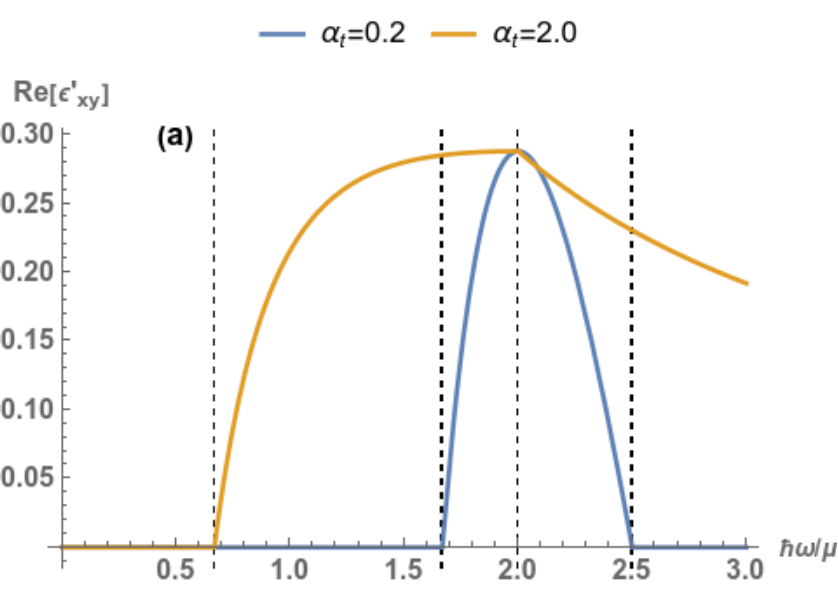} 
\includegraphics[scale=.2]{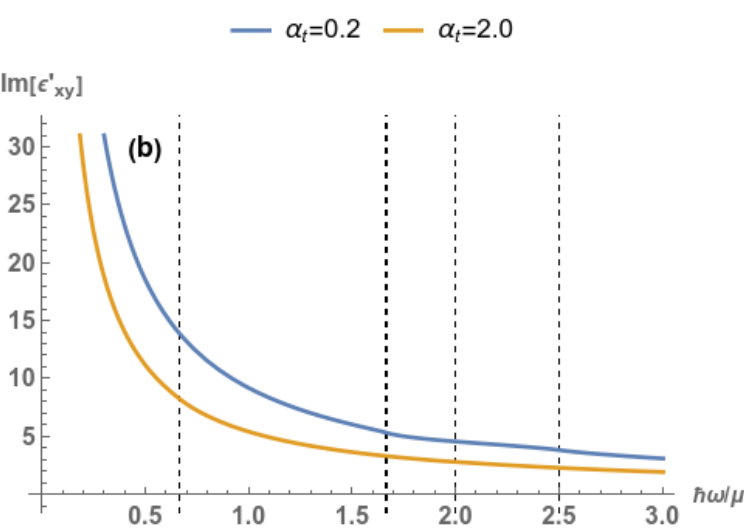} 
\includegraphics[scale=.2]{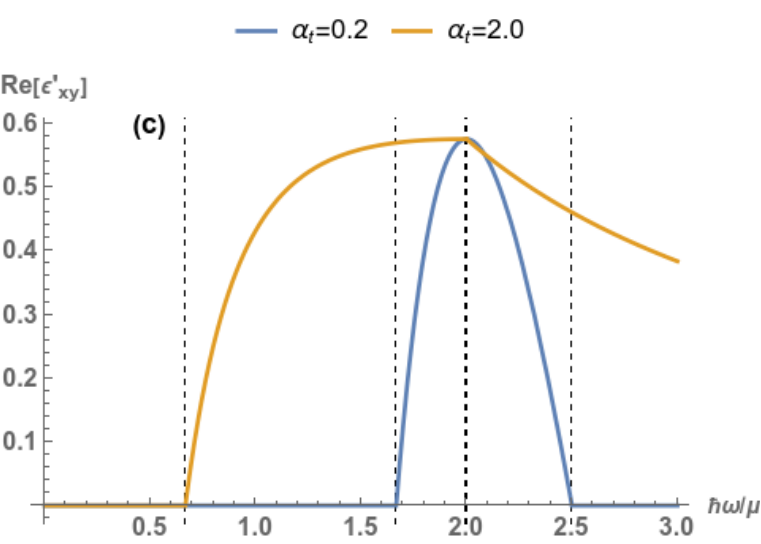}\quad
\includegraphics[scale=.2]{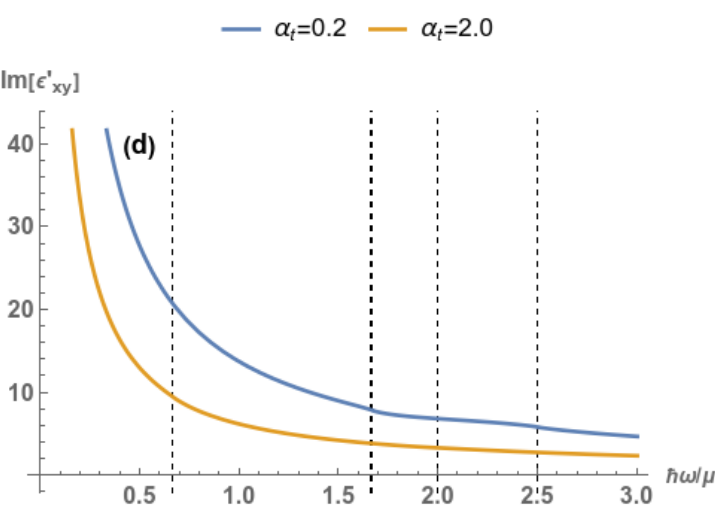}
\includegraphics[scale=.17]{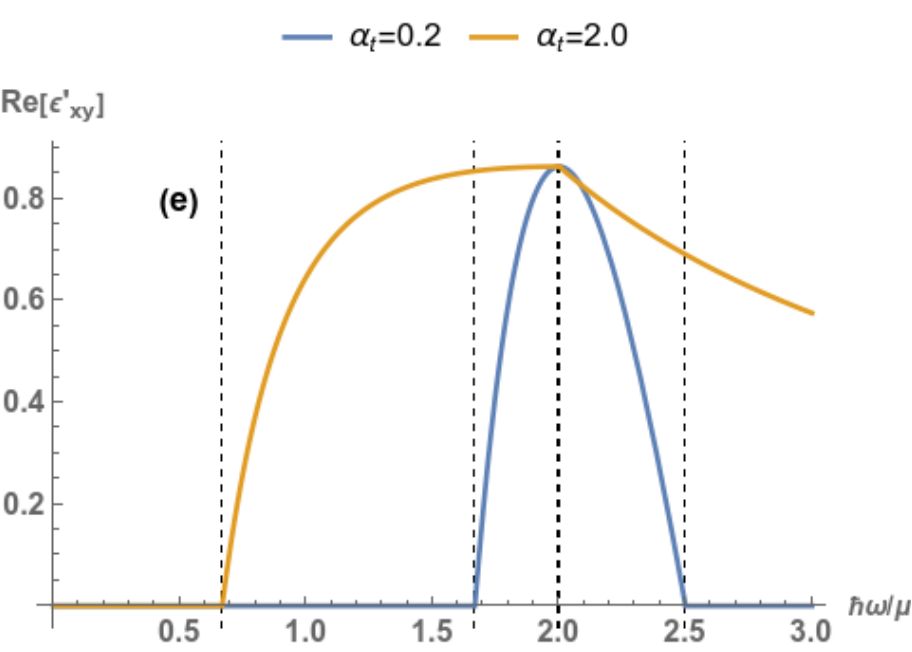}
\includegraphics[scale=.17]{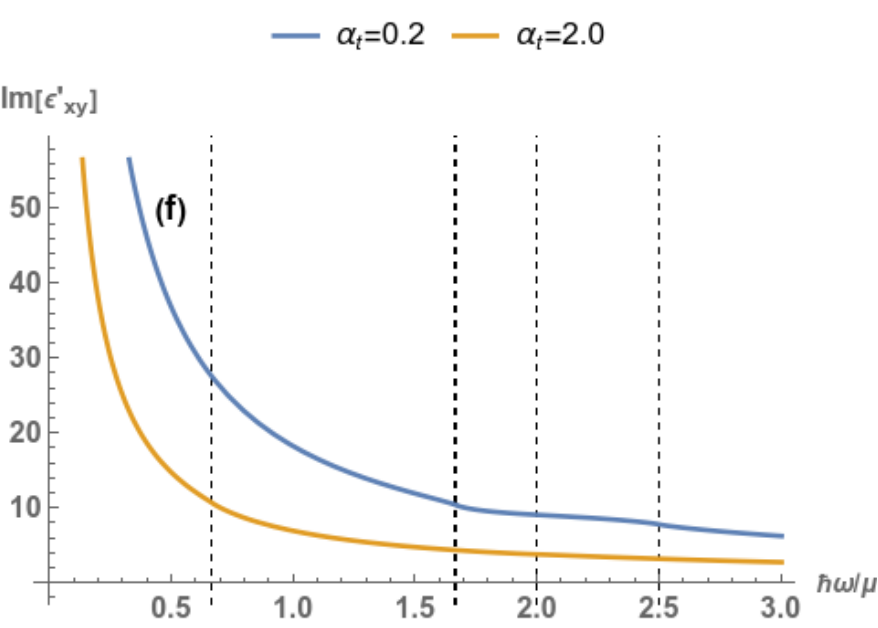}
\caption{(a),(b) show the real and imaginary parts of effective dielectric term $\varepsilon_{xy}$ for single WSM at  ($\alpha_t = 0.2,2.0 $). (c),(d) show the real and imaginary parts of effective dielectric term $\varepsilon_{xy}$ for double WSM at  ($\alpha_t = 0.2,2.0 $).(e),(f) show the real and imaginary parts of effective dielectric term $\varepsilon_{xy}$ for triple WSM at  ($\alpha_t = 0.2,2.0 $).}
\label{fig_dielectric_xy}
\end{figure}

\begin{eqnarray}
\textbf{k}_i=(0, k_i\sin\theta_i,k_i\cos\theta_i)\nonumber\\
\textbf{k}_r=(0, k_r\sin\theta_r,-k_r\cos\theta_r)\nonumber\\
\textbf{k}_t=(0, k_t\sin\theta_t,k_t\cos\theta_t)\nonumber
\end{eqnarray}

Consider an incident plane wave which is a superposition of the linear s- and p- polarized light,$\textbf{ E}_i$ = ( $E_i^s , E_i^p )^T $. The components of electric field $\textbf{E}$ are
\begin{eqnarray}
\textbf{E}_i=(E_i^s,E_i^p\cos\theta_i,-E_i^p\sin\theta_i)e^{i(\textbf{k}_i.\cdot\textbf{r}-\omega_i t)}\nonumber\\
\textbf{E}_r=(E_r^s,E_r^p\cos\theta_r,E_r^p\sin\theta_r)e^{i(\textbf{k}_r.\cdot\textbf{r}-\omega_i t)}\nonumber\\
\textbf{E}_i=(E_t^s,E_t^p\cos\theta_t,-E_t^p\sin\theta_t)e^{i(\textbf{k}_t.\cdot\textbf{r}-\omega_i t)}\nonumber
\end{eqnarray}

The magnetic field $\textbf{B}$ components are given by $\textbf{B}=(n/c)\hat{\textbf{k}}\times\textbf{E}$, with 'n' denoting the refractive index of the medium. The fields on the two sides of the WSM thin film are connected by the boundary conditions:
\begin{equation}
\textbf{E}_1^{\parallel}=\textbf{E}_2^{\parallel}, \ \ \hat{\textbf{n}}\times(\frac{\textbf{B}_1^{\parallel}}{\mu_1}-\frac{\textbf{B}_2^{\parallel}}{\mu_2})=\textbf{J}_s
\label{Maxwell}
\end{equation}
where $\textbf{E}_1=\textbf{E}_i+\textbf{E}_r$, $\textbf{E}_2=\textbf{E}_t$, $\textbf{B}_1=\textbf{B}_i+\textbf{B}_r$, $\textbf{B}_2=\textbf{B}_t$. The surface current density $\textbf{J}_s=\sigma^s\textbf{E}$, where the surface conductivity tensor $\sigma_{ij}^s=d \sigma_{ij}$ with the bulk $\sigma_{ij}$ for the m-WSM.\\

In general, the reflected electric field can be expressed as 
\begin{gather}
\begin{pmatrix}
E_r^p  \\ E_r^s
\end{pmatrix}
=
\begin{pmatrix}
r_{pp} & r_{ps}\\ r_{sp} & r_{ss}
\end{pmatrix}
\begin{pmatrix}
E_i^p \\ E_i^s
\end{pmatrix}
\label{matrix}
\end{gather}

Using this Eq.(\ref{matrix})) in Eq.(\ref{Maxwell}), we obtain the reflection coefficients as
\begin{eqnarray}
r_{sp}=\frac{E_r^s}{E_i^p}\vert_{E_i^s=0}=\frac{2 n_i}{c\mu_i}\frac{\sigma_{xy}^s\cos\theta_i \cos\theta_t}{(\sigma_{xy}^s\sigma_{yx}^s\cos\theta_i \cos\theta_t-\sigma_1^s\sigma_2^s)},
\label{reflection_1}
\end{eqnarray}
\begin{eqnarray}
r_{ss}&=&\frac{E_t^s}{E_i^s}\vert_{E_i^p=0}=-1\nonumber\\
&+&\frac{2 n_i}{c\mu_i}\frac{\sigma_{2}^s\cos\theta_i}{(\sigma_{1}^s\sigma_{2}^s-\sigma_1^s\sigma_2^s-\sigma_{xy}^s\sigma_{yx}^s\cos\theta_i \cos\theta_t)},
\end{eqnarray}
along with
\begin{eqnarray}
r_{pp}&=&\frac{E_t^p}{E_i^p}\vert_{E_i^s=0}=-1\nonumber\\
&+&\frac{2 n_i}{c\mu_i}\frac{\sigma_{1}^s\cos\theta_t}{(\sigma_{1}^s\sigma_{2}^s-\sigma_1^s\sigma_2^s-\sigma_{xy}^s\sigma_{yx}^s\cos\theta_i \cos\theta_t)},
\end{eqnarray}
and
\begin{eqnarray}
r_{sp}=\frac{E_r^s}{E_i^p}\vert_{E_i^s=0}=\frac{2 n_i}{c\mu_i}\frac{\sigma_{xy}^s\cos\theta_i \cos\theta_t}{(\sigma_{xy}^s\sigma_{yx}^s\cos\theta_i \cos\theta_t-\sigma_1^s\sigma_2^s)}
\label{reflection_4}
\end{eqnarray}

Here, we have defined $\sigma_1^s \equiv \frac{n_i\cos\theta_i}{c\mu_i}+\frac{n_t\cos\theta_t}{c\mu_t}+\sigma_{xx}^s$ and $\sigma_2^s \equiv \frac{n_i\cos\theta_t}{c\mu_i}+\frac{n_t\cos\theta_i}{c\mu_t}+\sigma_{yy}^s\cos\theta_i\cos\theta_t$.\\


For the case of normal incidence($n_i=1,n_t=n,\mu_1=\mu_2=\mu_0$ \cite{szechenyi2016transfer}), using Eq.(\ref{reflection_1}) to (\ref{reflection_4}) in Eq.(\ref{Kerr_angle}), we obtain

\begin{equation}
\chi_{kerr}^p=-\frac{\sigma_{xy}}{\sigma_{xx}}\Bigl[1+\frac{d \sigma_{xx}}{2c \epsilon_0}\Bigl(1+\frac{\sigma_{xy}^2}{\sigma_{xx}^2}\Bigl)\Bigl]^{-1}
\label{kerr_rot}
\end{equation}

This shows that that $\chi_{Kerr}^p \propto Q$, to lowest order in Q, as $\sigma_{xy}\propto Q$ and $\sigma_{xx}$ is independent of Q. 
In the limit of $d<<\lambda$, the d dependence of $\chi_{kerr}^p$ in thin film mWSMs is insignificant, therefore we can drop out the d dependence in Eq.(\ref{kerr_rot}) and the Kerr angle can be approximated as \cite{sonowal2019giant} 
\begin{equation}
\tan(2\Theta_{Kerr})=-2\frac{Re[\sigma_{xy}]Re[\sigma_{xx}]+Im[\sigma_{xy}]Im[\sigma_{xy}]}{\mid \sigma_{xx}\mid ^2-\mid \sigma_{xy}\mid ^2}
\end{equation}

The dependence of the Kerr angle of rotation for normal incidence on a freestanding tilted thin film m-WSM ($n_i=n_t\approx 1$ and $\theta_i=\theta_t$) is shown in Fig.(\ref{Kerr_angle}). The Kerr angle is of the order of a radian but indepndent of monopole charge J. Therefore, Kerr angle can't distinguish the m-WSMs  in thin film approximation.\\

\subsection{Incidence on surface with Fermi arc states ($\parallel Q$)}
In the case, light propagates($\hat{\textbf{n}}$) perpendicular to the z axis and the polarization of the incident electric field in the y-z(or x-z) plane. Consequently, the polarization rotating off diagonal reflection coefficients is
\begin{equation}
r_{sp}\propto \sigma_{zy}=0 \ or \ r_{sp}\propto \sigma_{zx}=0
\end{equation}

Consequently, there is no Kerr rotation. This offers an optical probe to distinguish the surfaces of a m-WSM which host Fermi arc surface states.\\

\section{Kerr Rotation in semi-infinite multi-WSMs}
We now focus our study on the polarization rotation due to a semi-infinite slab of a m-WSMs. The Maxwell equations are modified by an axoinic term inside m-WSMs. In TR broken m-WSms, the axionic field $\theta(\textbf{r},t)=2\textbf{Q}\cdot\textbf{r}$ leads to

\begin{eqnarray}
\textbf{D}=\epsilon\textbf{E}+c\epsilon_0\alpha_F\theta\textbf{B}/\pi\\
\textbf{H}=\textbf{B}/\mu_P-c\epsilon_0\alpha_F\theta\textbf{E}/\pi
\end{eqnarray}
Here $\epsilon$ and $\mu_P$ are dielectric and permeability tensors respectively.\\

Accordingly, wave equation for the electric field propagation is modified to be
\begin{eqnarray}
\nabla^2\textbf{E}-\nabla(\nabla\cdot\textbf{E})&=&\frac{1}{c^2\epsilon_0}\frac{\partial(\bf{\sigma} \cdot \bf E) }{\partial t}+\frac{\epsilon_b}{c^2}\frac{\partial^2 \bf{E}}{\partial t^2}\nonumber\\
&+&\frac{2\alpha_F}{\pi c}\textbf{Q}\times\frac{\partial \bf E}{\partial t}
\label{wave_eq}
\end{eqnarray}
where we have taken $\mu_P=\mu_0$ and $\epsilon_b$ is the static relative permittivity arising from bound charge polarization. The dispersion relation of electromagnetic wave is
\begin{equation}
k^2\textbf{E}-(\bf q.\textbf{E})\textbf{q}=\frac{\omega^2}{c^2}\tensor{\varepsilon'}\bf E
\end{equation}
Again, we need to consider two cases:(1) When the light is incident on the WSM surface hosting Fermi arc states($\hat{\bf n}\perp \bf Q$) and (2)When the light is incident on the WSM surface without Fermi arc states($\hat{n}\parallel \bf Q$). These two cases correspond to the Faraday and Voigt geometries discussed in the context of magnetic-optic effects in magnetic systems. In a TR broken m-WSMs, $\bf Q$ acts anologously to magnetization in a ferromagnet.

\subsection{Incidence on surface without Fermi arc states(magnetooptic(MO) polar Kerr effect)}
Let us consider a normal incidence of light on a surface without Fermi arcs, i.e., $\hat{n}\parallel \textbf{ Q}= Q \hat{z}$. 
Intrduce $k=n\omega/c$, the wave propagation has following matrix equation \cite{visnovsky2018optics}
\begin{gather}
n^2 \begin{pmatrix}
E_x  \\ E_y \\ 0
\end{pmatrix}
=
\begin{pmatrix}
\epsilon'_{xx} & \epsilon'_{xy} & 0 \\ -\epsilon'_{xy} & \epsilon'_{yy} & 0 \\0 & 0 &\epsilon'_{zz}
\end{pmatrix}
\begin{pmatrix}
E_x \\ E_y \\E_z
\end{pmatrix}
\label{wave_matrix}
\end{gather}
 
Here, n represent the complex refractive index and $\epsilon'_{ij}$ is the complex dielectric function including the axion terms. 
The dielectric tensor $\varepsilon_{ij}$ are given by
\begin{equation}
\varepsilon'_{ij}=\varepsilon_b \delta_{ij}+\frac{i}{\epsilon_0\omega}\sigma_{ij}
\end{equation}
The digonal dielectric elements are
\begin{eqnarray}
\epsilon'_{ii}=\epsilon_b +\frac{i}{\omega \epsilon_0}\sigma_{ii}, \ i=x,y,z \  
\end{eqnarray}

The off-diagonal element is 
\begin{equation}
\epsilon'_{xy}=\frac{i}{\omega \epsilon_0}\sigma_{xy}+\frac{2 i \alpha_F c}{\pi}\frac{Q}{\omega}=\frac{i}{\omega \epsilon_0}(\sigma_{xy}+\sigma_Q)
\end{equation}
Here $\sigma_{ij} $ denotes complex optical conductivities.\\
The real and imaginary parts of all componens of modified dielectric tensor are plotted in Fig.(\ref{fig_dielectric_xx}), Fig.(\ref{fig_dielectric_zz}), Fig.(\ref{fig_dielectric_xy}). Eq.(\ref{wave_matrix}) have the following nontrivial solution for the electromagnetic fields 
\begin{equation}
n^2_+ =\epsilon'_{xx}+ i\epsilon'_{xy} \ and \ n^2_- =\epsilon'_{xx}- i\epsilon'_{xy} 
\end{equation}

Here, $n_{+}$  and  $n_- $ are refractive indices of the left and right circularly polarized modes in the m-WSMs \cite{gorbar2017anomalous}.\\

The Fresnel reflection coefficients corresponding to these modes are
\begin{equation}
r_{\pm}=(1-n_{\pm})/(1+n_{\pm})
\end{equation}

The dimensionless constant $\chi_{PKE}$ is defined as
\begin{equation}
\chi_{PKE}=i \frac{r_{+}-r_{-}}{r_{+} + r_{-}}=i\frac{n_{+}-n_{-}}{n_+ n_{+}-1}
\end{equation}
 
\begin{equation}
\chi_{PKE}=i \frac{(\sqrt{\epsilon'_{xx}+i\epsilon'_{xy}}-\sqrt{\epsilon'_{xx}-i\epsilon'_{xy}})}{\sqrt{\epsilon'_{xx}-i\epsilon'_{xy}}\sqrt{\epsilon'_{xx}+i\epsilon'_{xy}}-1}
\end{equation} 
\begin{figure} [t] 
\includegraphics[scale=.15]{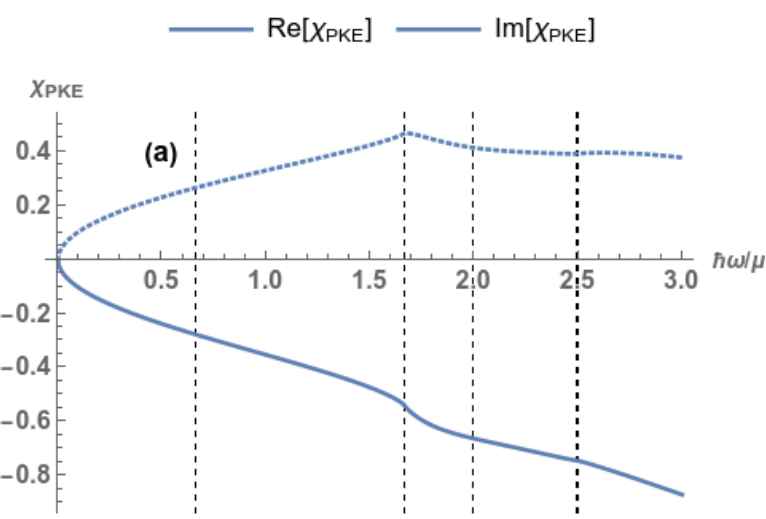} 
\includegraphics[scale=.15]{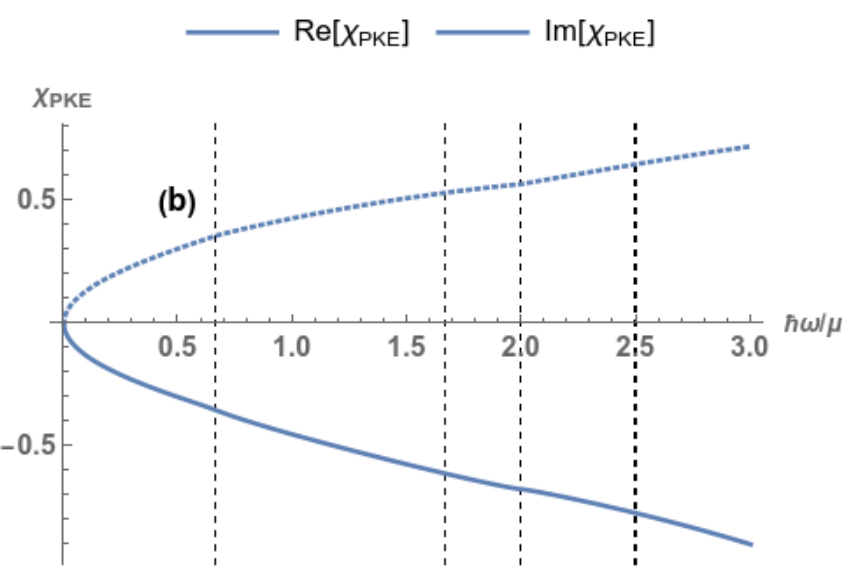} 
\includegraphics[scale=.15]{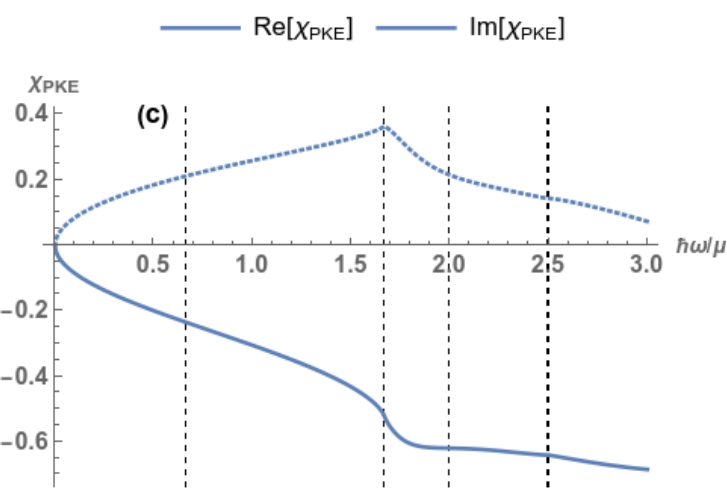}\quad
\includegraphics[scale=.15]{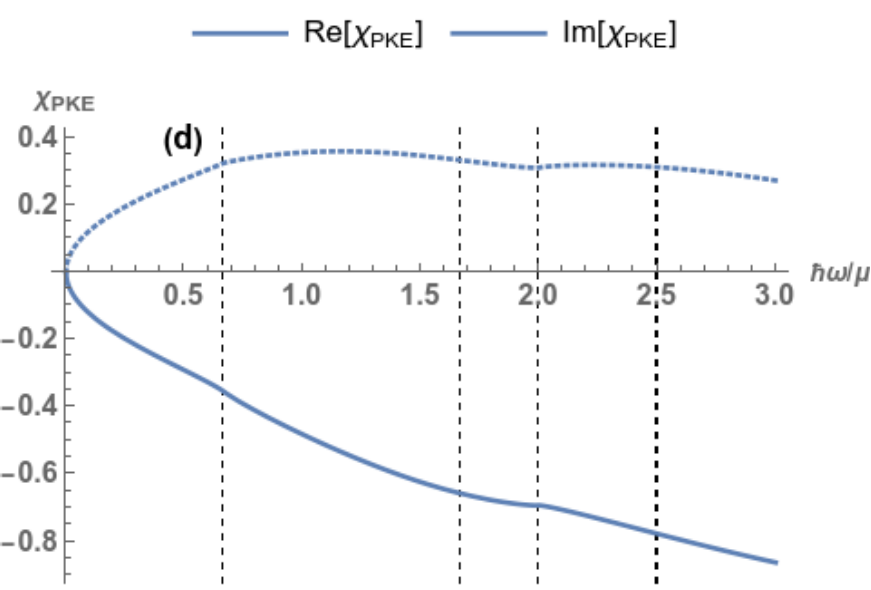}
\includegraphics[scale=.15]{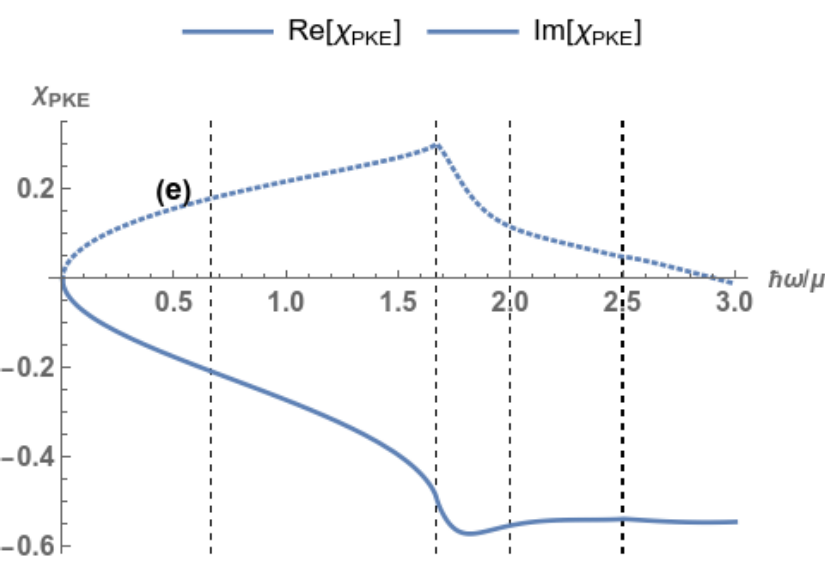}
\includegraphics[scale=.15]{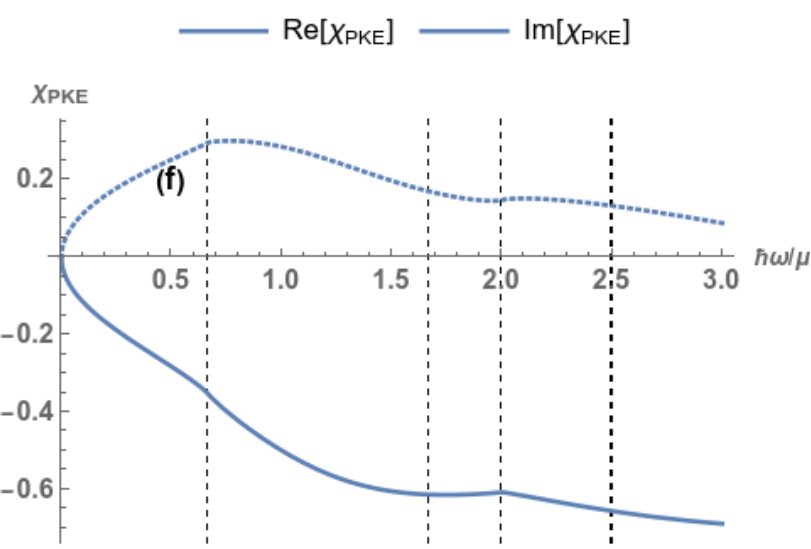}
\caption{(a),(b) show real(blue) and imaginary(dotted) parts of $\chi_{PKE}$ for single WSM at $\alpha_t = 0.2$, $\alpha_t = 2.0$ respectively.(c),(d) show real(blue) and imaginary(dotted) parts of $\chi_{PKE}$ for double WSM at $\alpha_t = 0.2$, $\alpha_t = 2.0$ respectively.(e),(f) show real(blue) and imaginary(dotted) parts of $\chi_{PKE}$ for triple WSM at $\alpha_t = 0.2$, $\alpha_t = 2.0$ respectively. } 
\label{fig_pke}
\end{figure}

\begin{figure} [t] 
\includegraphics[scale=.2]{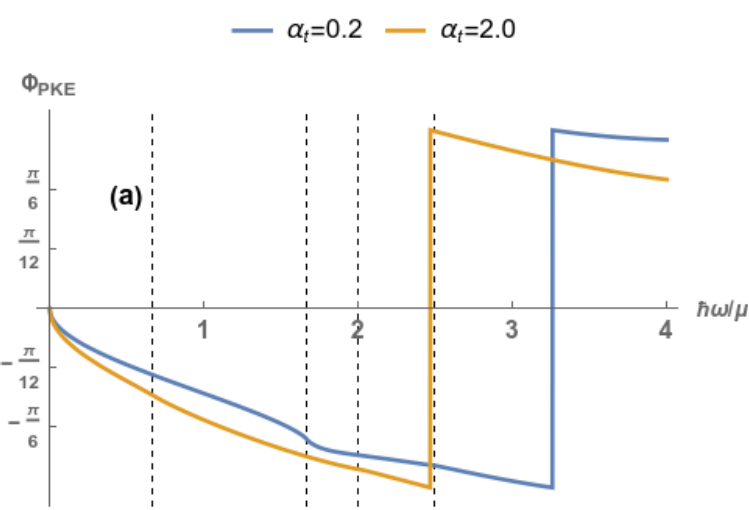} 
\includegraphics[scale=.16]{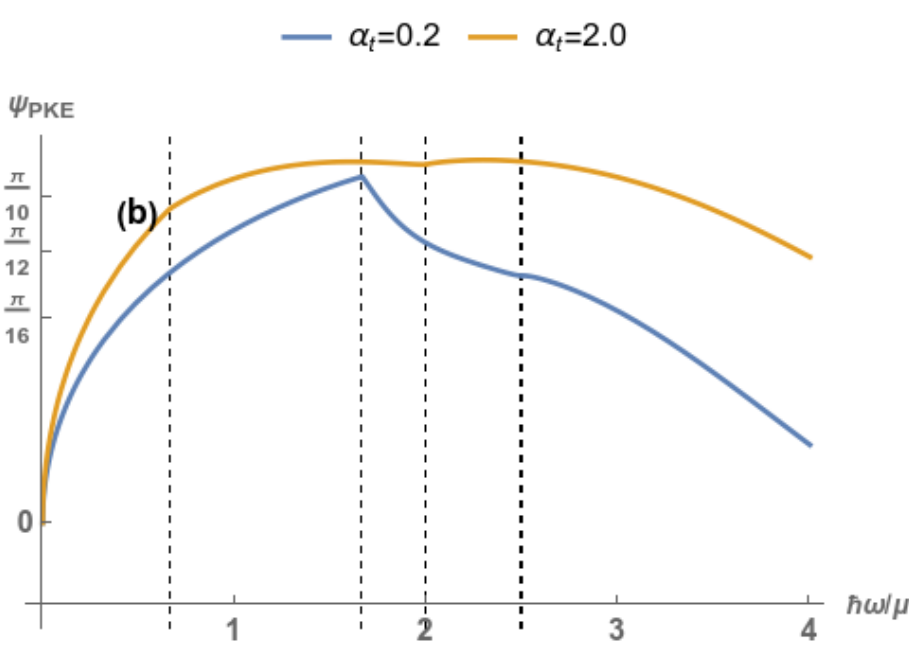} 
\includegraphics[scale=.18]{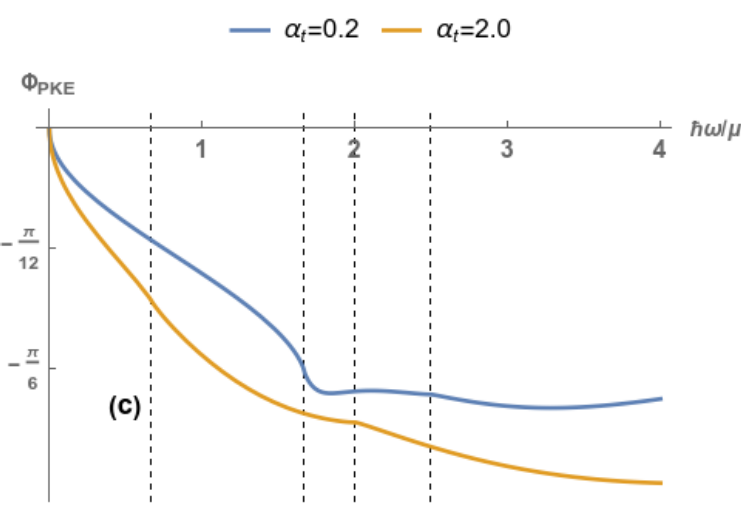}\quad
\includegraphics[scale=.16]{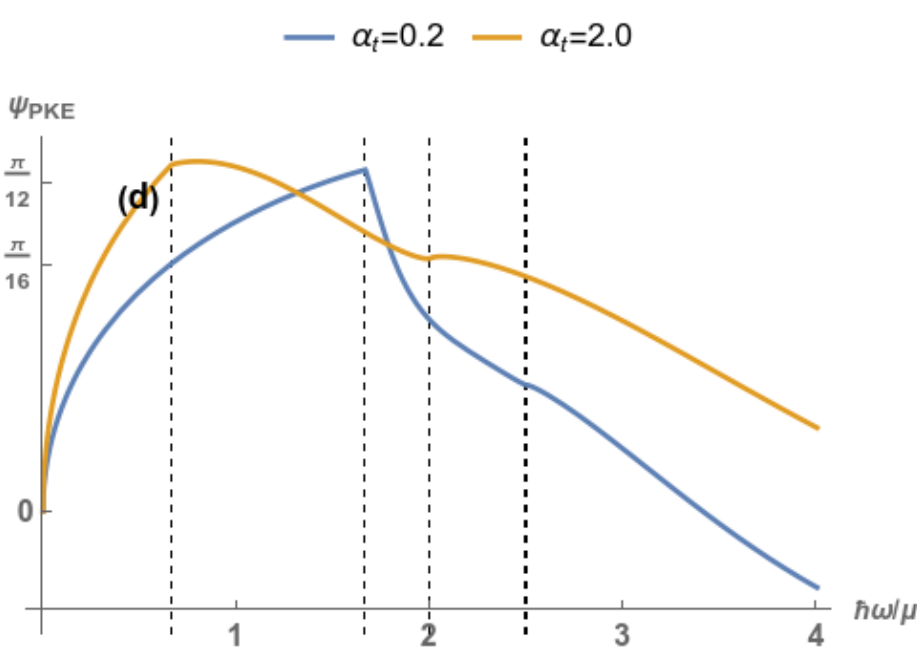}
\includegraphics[scale=.17]{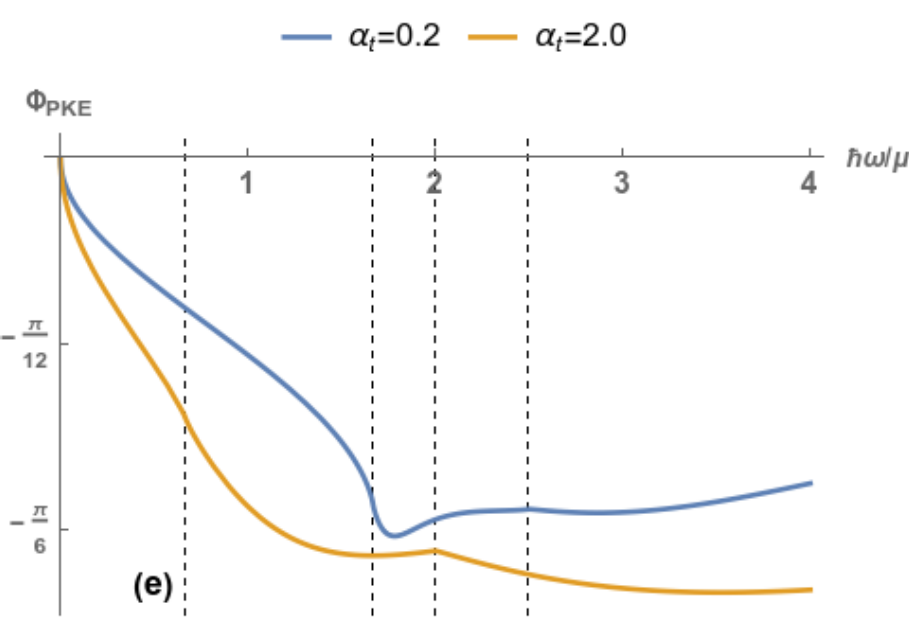}
\includegraphics[scale=.17]{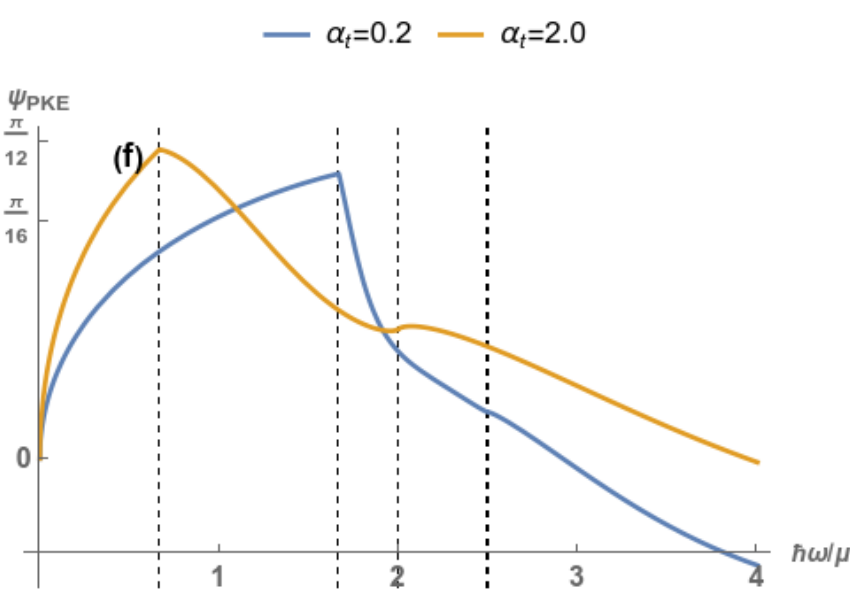}
\caption{(a),(b) show the Kerr rotation and Kerr ellipticity respectively for single WSM at $\alpha_t = 0.2,2.0$. (c),(d) show the Kerr rotation and Kerr ellipticity respectively for double WSM at $\alpha_t = 0.2,2.0$. (e),(f) show the Kerr rotation and Kerr ellipticity respectively for triple WSM at $\alpha_t = 0.2,2.0$.} 
\label{fig_kerr_angle_ellip}
\end{figure}
The MO polar Kerr rotation and corresponding MO polar Kerr ellipticity are
\begin{eqnarray}
\tan 2\Theta_{PKE}= \frac{2\mathcal{R}(\chi_{PKE})}{1-\mid \chi_{PKE} \mid^2},\nonumber\\
\sin 2\Psi_{PKE}=\frac{2\mathcal{I}(\chi_{PKE})}{1+\mid \chi_{PKE} \mid^2}
\end{eqnarray}

The MO polar Kerr rotation and corresponding MO polar Kerr ellipticity are plotted in Fig.(\ref{fig_pke}) and Fig.(\ref{fig_kerr_angle_ellip}). Thus, Kerr rotation in this configuration depends on monopole charge J.

\subsection{Incidence on surface with Fermi arc states(Voigt effect)}
Let us consider the propagation of the wave in x-direction. In this case, the wave equation becomes \cite{visnovsky2018optics}
\begin{gather}
n^2 \begin{pmatrix}
E_x  \\ E_y \\ 0
\end{pmatrix}
=
\begin{pmatrix}
\epsilon'_{xx} & \epsilon'_{xy} & 0 \\ -\epsilon'_{xy} & \epsilon'_{yy} & 0 \\0 & 0 &\epsilon'_{zz}
\end{pmatrix}
\begin{pmatrix}
E_x \\ E_y \\E_z
\end{pmatrix}
\label{wave_matrix2}
\end{gather}

The two solutions are
\begin{equation}
n_{\parallel}^2=\epsilon'_{zz} \ and \ n_{\perp}^2=\epsilon'_{yy}-\frac{\epsilon'^2_{xy}}{\epsilon'_{xx}}
\end{equation}
\begin{figure} [t] 
\includegraphics[scale=.2]{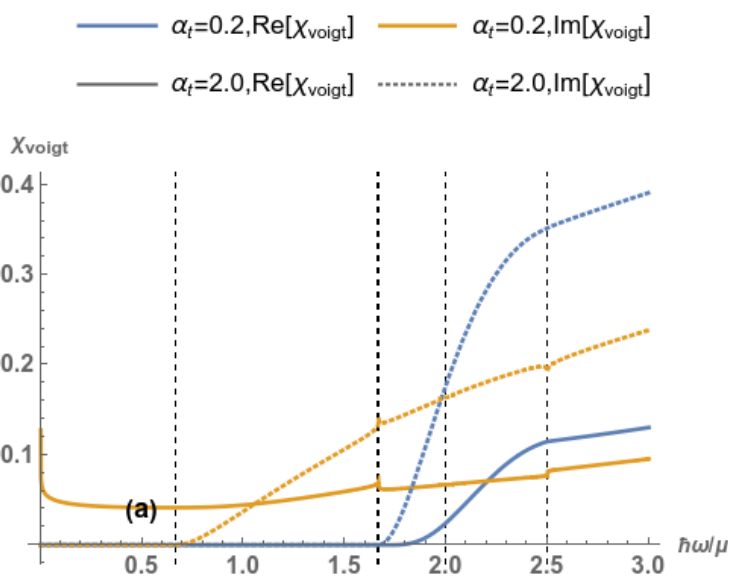} 
\includegraphics[scale=.2]{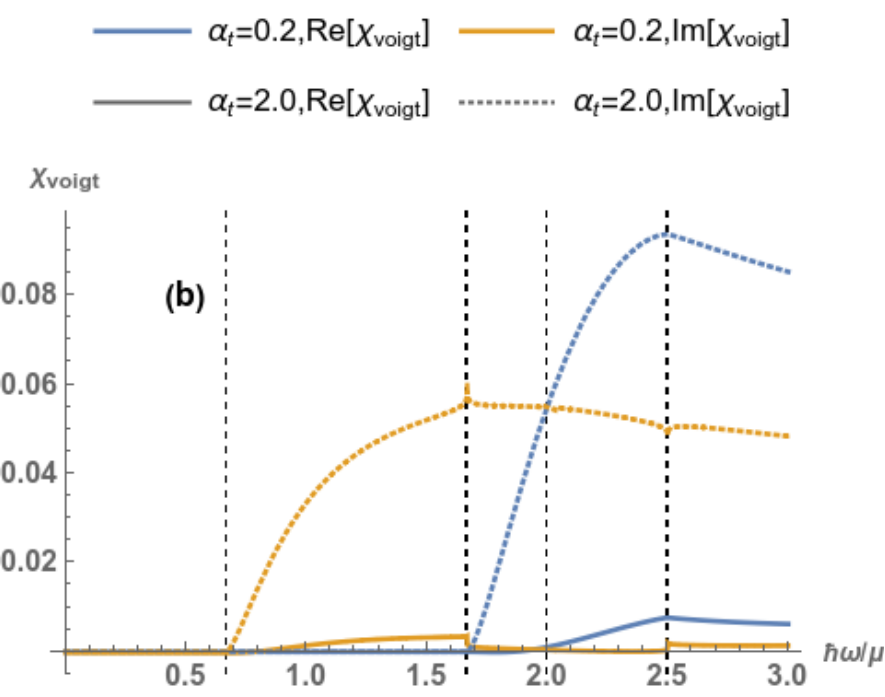}\quad
\includegraphics[scale=.22]{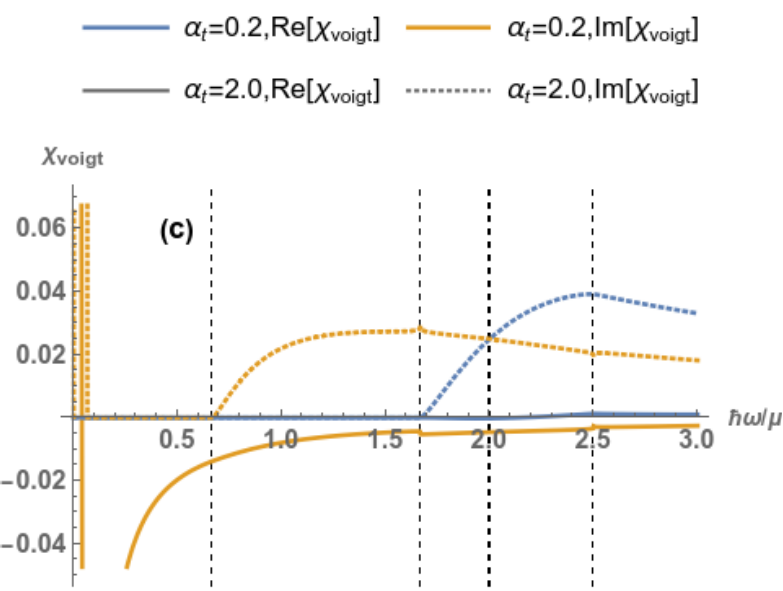}
\caption{(a) show the real and imaginary parts of $\chi_{Voigt}$ for single-WSM at $\alpha_t=0.2,2.0$. (b) show the real and imaginary parts of $\chi_{Voigt}$ for double-WSM at $\alpha_t=0.2,2.0$. (c) show the real and imaginary parts of $\chi_{Voigt}$ for triple-WSM at $\alpha_t=0.2,2.0$.} 
\label{fig_chi_voigt}
\end{figure}
Let $\Theta_0$ be an arbitrary angle with respect to $\bf Q$, the electric field components are 
\begin{equation}
E_{\parallel}=E_I \cos(\Theta_0),\ \ E_{\perp}=E_I \sin(\Theta_0),
\end{equation}
Then the beam becomes elliptically polarized upon reflection from the mWSM slab. The correponding dimensionless quantity $\chi_{Voigt}$ is defined as

\begin{equation}
\chi_{Voigt}=\frac{r_{\perp}E_{\perp}}{r_{\parallel}E_{\parallel}}=\frac{r_{\perp}}{r_{\parallel}}\tan(\Theta_0)
\end{equation}
where
\begin{equation}
r_{\parallel/\perp}=\frac{1-n_{\parallel/\perp}}{1+n_{\parallel/\perp}}
\end{equation}

Therefore
\begin{equation}
\chi_{Voigt}=\frac{(1+\sqrt{\epsilon'_{xx}-\frac{\epsilon'_{xy}}{\epsilon'_{xx}}})(\sqrt{\epsilon'_{zz}}-1)}{(-1+\sqrt{\epsilon'_{xx}-\frac{\epsilon'_{xy}}{\epsilon'_{xx}}})(\sqrt{\epsilon'_{zz}}+1)}\tan(\Theta_0)
\end{equation}

Therefore, the relative polarization rotation is
\begin{equation}
\delta_{Voigt}=\Theta_{Voigt}-\Theta_0
\end{equation}
The rotation($\Theta_{Voigt}$) and corresponding ellipticity($\Psi_{Voigt}$) are given by
\begin{eqnarray}
\tan 2\Theta_{Voigt}= \frac{2\mathcal{R}(\chi_{Voigt})}{1-\mid \chi_{Voigt} \mid^2},\nonumber\\
\sin 2\Psi_{Voigt}=\frac{2\mathcal{I}(\chi_{Voigt})}{1+\mid \chi_{Voigt} \mid^2}
\end{eqnarray}

\begin{figure} [t] 
\includegraphics[scale=.18]{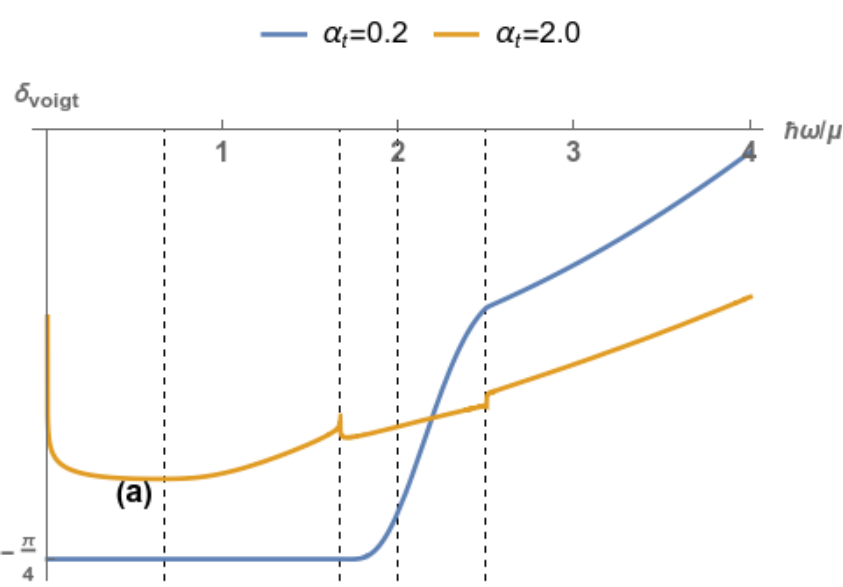} 
\includegraphics[scale=.18]{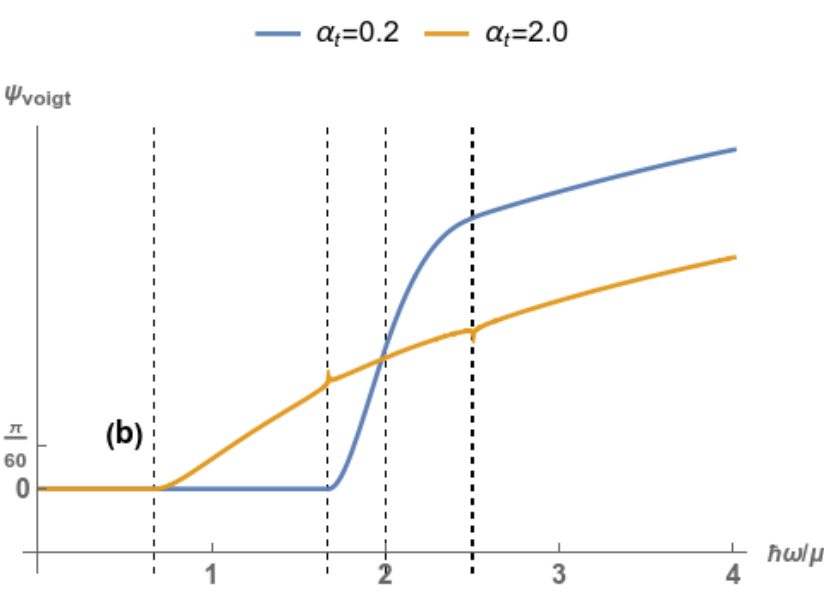} 
\includegraphics[scale=.18]{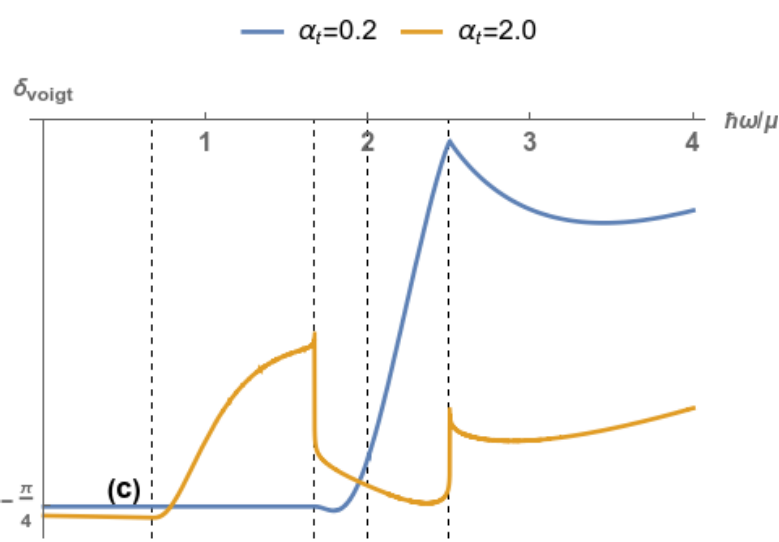}\quad
\includegraphics[scale=.18]{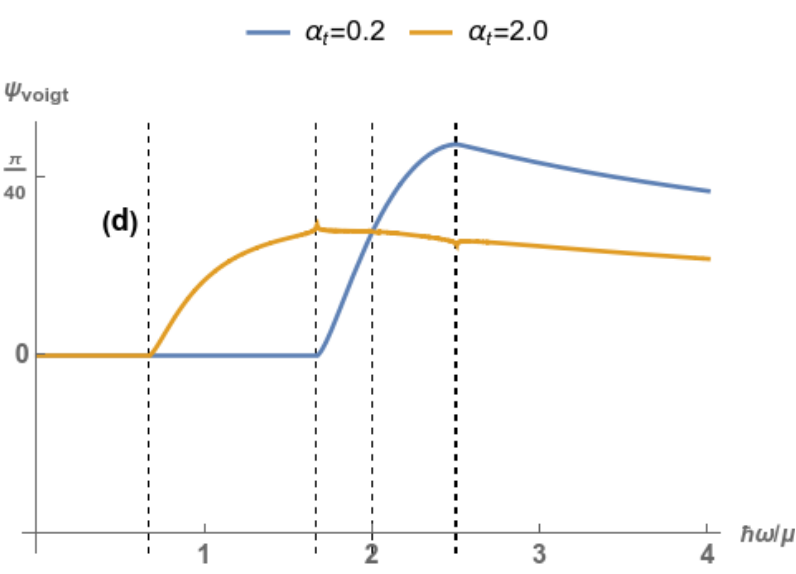}
\includegraphics[scale=.18]{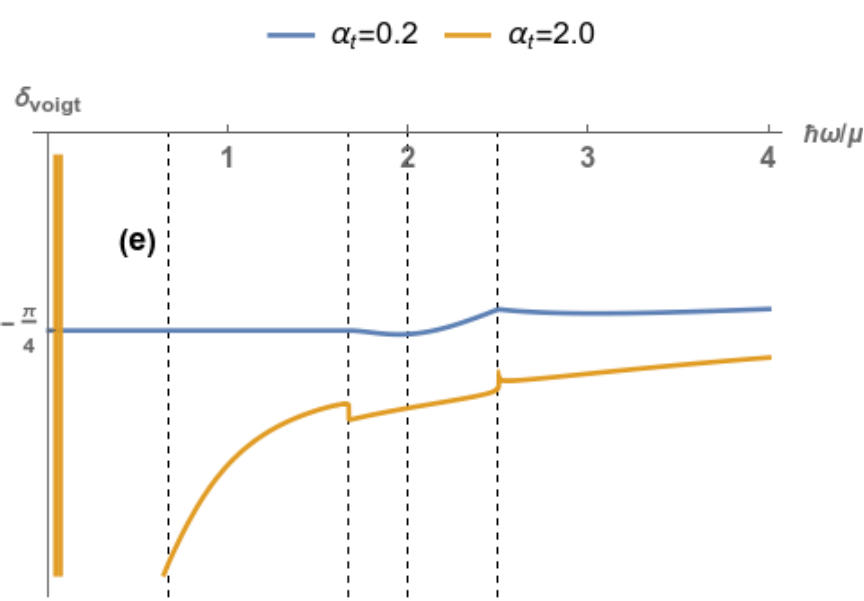}
\includegraphics[scale=.18]{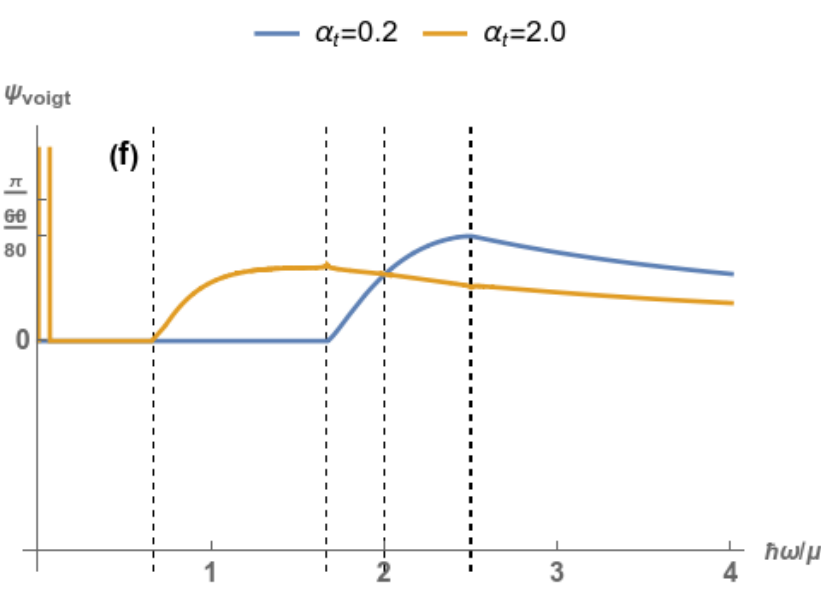}
\caption{(a),(b) show the $\delta_{Voigt}$ and $\Psi_{Voigt}$ for single-WSM at $\alpha_t=0.2,2.0$. (c),(d) show the $\delta_{Voigt}$ and $\Psi_{Voigt}$ for double-WSM at $\alpha_t=0.2,2.0$. (a),(b) show the $\delta_{Voigt}$ and $\Psi_{Voigt}$ for triple-WSM at $\alpha_t=0.2,2.0$.} 
\label{fig_kerr_voigt}
\end{figure}

The Voigt rotation and corresponding Voigt ellipticity are plotted in Fig.(\ref{fig_chi_voigt}) and Fig.(\ref{fig_kerr_voigt}). Thus, the Kerr rotaion in this configuartin also depends on monopole charge and systems parameters $v_f$,$\alpha_2$,$\alpha_3$.

\section{Conclusions}

We have calculated the full optical conductivity tensor for a mWSMs, which in turn has made it possible to determine the Kerr rotation. We have found that the tilt
direction plays a dominating role in the determination of the conductivity. The diagonal components $\sigma_{xx/yy}=J\omega$, $\sigma_{zz}=\omega^{\frac{2}{J}-1}/J$, while the off-diagonal Hall term $\sigma_{xy} \propto J Q$ , where Q denotes the Weyl node separation (in the z-direction) and $\sigma_{yz}=\sigma_{zx}=0$. 
We have shown that a sizeable Kerr rotation is possible in both ultra-thin film and bulk m-WSMs and it depends on various system parameters. The Kerr angle is independent of monopole charge J for light incident ultra-thin fil of m-WSMs without the Fermi arc states (parallel to the node separation). In contrast, the Kerr angle depends on monopole charge in bulk m-WSMs in both Polar Kerr and Viogt configurations.

\section{Acknowledgements}
We thank Debanand Sa for fruitful discussions.  
\bibliographystyle{elsarticle-num}
\bibliography{References,bib}

\end{document}